\documentclass[traditabstract]{aa} 
\usepackage{graphicx}
\usepackage{txfonts}
\usepackage{rotfloat}
\usepackage{color}
\usepackage{float}
\usepackage{wrapfig}
\usepackage{textcomp}
\usepackage{amssymb}
\usepackage{graphicx}
\usepackage{subscript}
\usepackage{wasysym}
\usepackage{subfig}

\begin{document}
\title{Spectroscopic diagnostics of dust formation and evolution in classical nova ejecta}
\author{Steven N. Shore\inst{1,2}\and N. Paul Kuin \inst{3}\and Elena Mason\inst{4}\and Ivan De Gennaro Aquino\inst{5} }
\institute{
Dipartimento di Fisica "Enrico Fermi'', Universit\`a di Pisa; \email{steven.neil.shore@unipi.it}
\and 
INFN- Sezione Pisa, largo B. Pontecorvo 3, I-56127 Pisa, Italy
\and
Mullard Space Science Laboratory, University College London, Holmbury St Mary, Dorking, ,Surrey RH5 6NT,  UK
\and
INAF-OATS, Via G.B. Tiepolo, 11, I-34143, Trieste, Italy 
\and
Hamburger Sternwarte, Gojenbergsweg 112, D-21029 Hamburg, Germany
 }

\date{First version: 10/4/2018, Revised  02/06/2018 ; accepted ---}

\abstract{A fraction of classical novae  form dust during the early stages of their outbursts.   The classical CO nova V5668 Sgr (Nova Sgr. 2015b) underwent a deep photometric minimum about 100 days after outburst that was covered across the spectrum.  A similar event was observed for an earlier CO nova, V705 Cas (Nova Cas 1993) and a less optically significant event for the more recent CO nova V339  Del (Nova Del 2013).   This study provides a ``compare and contrast'' of these events to better understand the very dynamical event of dust formation.  We show  the effect of  dust formation on multiwavelength high resolution line profiles in the interval 1200\AA\ - 9200\AA\  using a biconical ballistic structure that has been applied in our previous studies of the ejecta.  We find that both V5668 Sgr and V339 Del  can be modeled using a grey opacity for the dust, indicating fairly large grains ($\gtrsim$ 0.1$\mu$)  and that the persistent asymmetries of the line profiles in late time spectra, up to 650 days after the event for V5668 Sgr and 866 days for V339 Del, point to the survival of the dust well into the transparent, nebular stage of the ejecta evolution.  This is a general method for assessing the properties of dust forming novae well after the infrared is completely transparent in the ejecta.} 


  \titlerunning{Multiwavelength high resolution spectroscopic signatures of dust in novae} \authorrunning{Shore, Kuin, Mason, De Gennero Aquino}
     \maketitle


\section{Introduction}

Dust formation in expanding ejecta and stellar atmospheres has been an open question for many decades.   For classical novae, the possibility of some sort of grain production was first raised in the 1930s, connected with the early observations and interpretation of the photometric development of DQ Her 1934  (see Payne-Gaposchkin 1957).  In this paper we study the recent deep minimum dust former V5668 Sgr as a paradigmatic case, employing multiwavelength spectroscopy using high spectral resolution.  We show that such an approach provides unique information based on line profile development on the evolution and distribution of the dust during and after the grain formation stage.   Specifically, we show how it is possible to consistently simulate the line profile changes during and after the event within a self-consistent model, a single ejection ballistically expanding shell, to derive information about the dust properties and survival.   We present evidence that the dust survives the interval of intense XR irradiation during the optically thin supersoft X-ray source stage following the explosion, and that the grains are therefore likely to survive until mixed with the interstellar gas at late times.  Finally, using the evidence from line profile and photometry, we connect the dust forming events in several recent CO novae with that in V5668 Sgr, especially V339 Del and V1369 Cen, and revisit the dust forming episodes in several historical classical novae.

\section{Modeling the effect of dust on the profile}

At present there is no ``first principles" theory of dust formation in nova ejecta.  A number of proposals, none definitive, have been presented over the years for the formation and growth of the grains.  Each depends on a specific scenario, such as chemistry (Rawlings 1988), kinetic agglomeration and photoionization processing (Shore \& Gehrz 2004), and shock driven chemistry (Derdzinski et al. 2017).   All have in common that the matter must be sufficiently kinetically cold to allow grain growth but when this occurs is a disputed matter.   Equally contentious is the fate of the grains.  The DQ Her-like events have been the optical indication of dust formation and/or growth in the post-maximum stage, but the signature of the process is written in the infrared (e.g., Evans \& Gehrz 2012). This includes continuum and PAH emission features in the near IR and the broad silicate-related emission features at 10-20$\mu$.  The decrease in the IR luminosity is usually interpreted as grain destruction when the ejecta become sufficiently transparent that ultraviolet and X-ray illumination by the remnant hot white dwarf reaches the part of the ejecta harboring the newly formed dust.  We will show in this section that there is a completely independent means for understanding the growth and destruction process in the ejecta using spectroscopic observations of systematic changes in the emission line profiles arising from the expanding ejecta.

In light of this theoretical uncertainty, we approach the problem phenomenologically.  We assume that dust forms in an intermediate part of the ejecta where the temperature has fallen below some critical value (e.g. the Debye temperature) and that has a sufficient density that the process is to proceed.   The ejecta are assumed to be ballistic (e.g. Shore 2013,  Mason et al. 2018 and references therein) with an aspherical (bipolar) geometry.  While the latter is a quite general feature of classical nova ejecta, it is not essential for the argument.  In contrast, however, the ballistic nature of the event is basic.   It provides a unique correspondence between a location in the line profile in radial velocity, relative to the observer, and radial position within the ejecta since the expansion velocity is linear relative to distance $r$ from the white dwarf.   The geometric parameters are  (i) the radial thickness of the ejecta  scaled to the maximum radius $\Delta r/R(t) = \Delta r/(v_{max}t)$, (ii)  the maximum (ballistic) expansion velocity $v_{max}$; (iii) the opening half-angle of the cone $\theta$;  and (iv) the inclination, $i$, of the axis to the line of sight.   Since the mass of the ejecta is constant, the density varies as $\rho \sim r^{-3}$ ,  $\Delta r/R(t)$ is constant from self-similarity, and the dust optical depth varies with time as $\tau_d \sim t^{-2}$ (assuming no changes in grain properties, see below).  For the emission lines, we assume they are optically thin, formed under isothermal conditions, and form with an emissivity law $\epsilon \sim \rho^{n}$ where $n$ is either 1 or 2, appropriate for collisional excitation and recombination.  For the emission lines, a set of uniform random numbers is generated
to choose the locations of parcels of gas within the ejecta.  The
observed profile is then formed by integrating through the cone
whose inclination to the line of sight is $i$, and the profile is
binned in radial velocity, in most cases $\Delta v_{rad}=$ 30 km s$^{-1}$.  In most of the
simulations, to mimic the fragmented structure of the observed
profiles, a comparatively small number of packets are formed,
usually a (few)$\times 10^4$, hence the jagged appearance in the
profiles we present here.   It has been noted in the observations
that there is considerable fine structure on the emission line and since we aim to capture the range of geometric parameters (e.g.
solid angle subtended by the ejecta) rather than fit the individual
lines, these free parameters are varied until a good match is
obtained.  To emphasize, we aim at  constraints on the geometry, density structures, and dynamics and not at deriving specific parameter values by fitting the line profiles.  If the grains appear in some interval $[r_{min},r_{max}]$, both scaled to $R$, this corresponds to some velocity interval $\pm[v_{rad,min},v_{rad,max}]$ so the obscuration can be easily included in the previously employed Monte Carlo modeling (e.g. De Gennaro Aquino et al. 2014, Mason et al. 2018) by inserting a grey absorber in the ejecta in the selected radial interval since the emitting gas and the dust coexist.    

The effect of the dust absorption is to shadow a portion of the ejecta.   Both lobes must be seen through the screen (which is also situated within the line forming gas) but with the receding portion being the more heavily absorbed.   Whatever the obscuration of the approaching lobe, and that will be mainly at the lower velocities, the receding part of the profile will always be  weaker.  Two simple consequences are that the profile will be asymmetric and, even at relatively low resolution, the mean emission will be blueshifted.  We note that this is contrary to the effect of {\it line} opacity, as in a P Cyg profile.  There is no fine tuning here, the dust opacity, being continuous, is grey across the line profile and the extinction depends only on the line of sight column density of the dust.   How the line profile changes because of an increasingly larger dust column density is shown in Fig 1. \footnote{{\bf It has been known for some time, since the original realization by Lucy et al. (1989) for SN 1987A and Pozzo et al. (2004) for a more general study,  that supernovae also show profile asymmetries that indicate dust formation, for instance, Bevan \& Barlow (2016) and references therein.  The nova case is less ambiguous, in large part due to the relative simplicity of the geometry and dynamics, and also because the ejecta masses are so small.  Another reason is that the central source remains alight after the explosion and the ejecta are passive screens, unlike  supernovae.}}

 \begin{figure}
    \includegraphics[width=9.5cm]{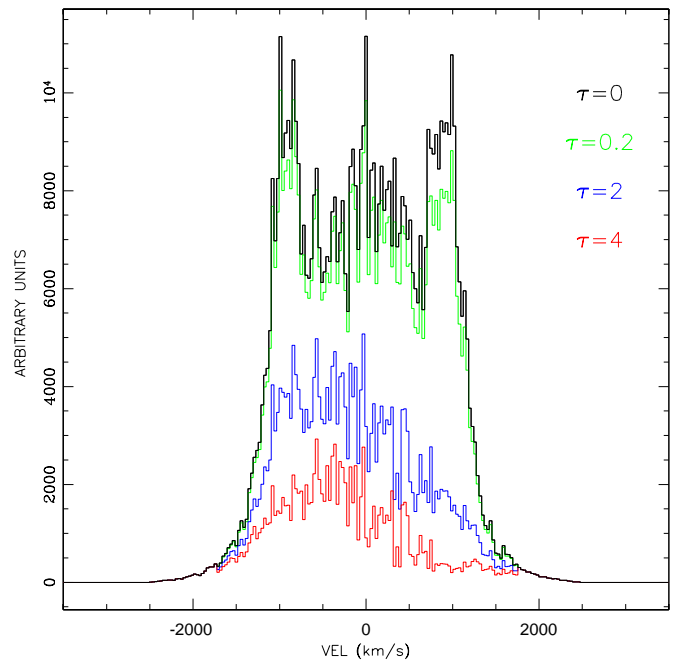}
      \caption{Effect of the dust on the line profile. The adopted ejecta geometry and dust distribution are invariant and only the dust optical depth, $\tau_d$, is varied (values reported on the figure itself). The unabsorbed line profile was produced assuming a biconical ejecta having cone opening half-angle $\theta$=70$^o$ and thickness $\Delta r/R=0.6$, viewed from an inclination, $i=40^o$. The dust was assumed to be in the inner half of the ejecta. }     
         \label{f00}
   \end{figure}

The dust is assumed to have no effect on the intrinsic line formation, we treat it as an absorbing screen, and for simplicity we ignore the effects of scattering on the NLTE line formation and photometry.   We will, however, return to this point in the discussion (below) since it too has a consequence for the connection between the photometric and spectroscopic signatures for nonspherical ejecta.  Even in cases where the line of sight barely intersects the inner ejecta because of the inclination of the ejecta, the line profiles will be asymmetric.   Dust forming novae that do not show a deep photometric minimum can still be identified by their spectroscopic development.   The dust optical depth is derived from the geometric factors alone.  The change in the dust luminosity and  temperature can also be derived immediately from the ballistic scaling of $\tau_d$, although the value of the temperature and luminosity at any stage requires additional information  (see discussion, below).

 \begin{figure}
    \includegraphics[width=9.5cm]{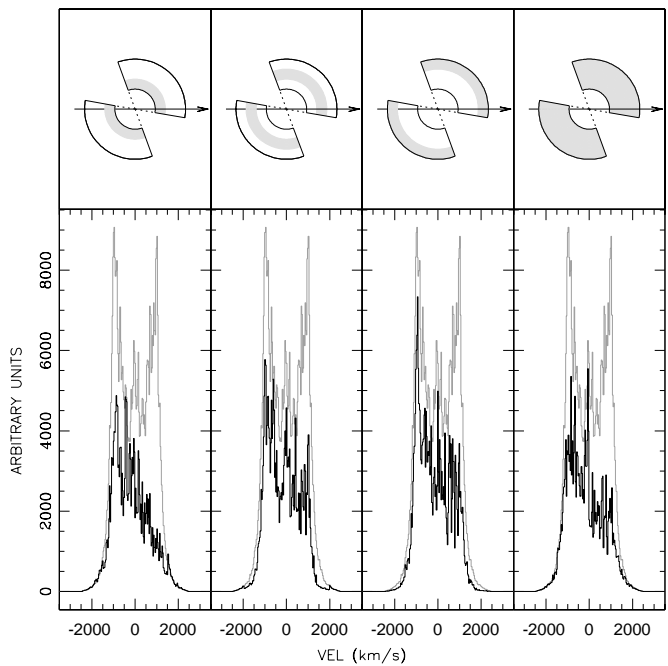}
      \caption{Effect on the line profile by the location of the dust within the ejecta. The top row of panels show the adopted ejecta geometry (a bicone having opening half-angle $\theta$=60$^o$ and thickness 0.6 viewed from an inclination of $i$=40$^o$) and the dust location (gray shaded area). The bottom row of panels show the resulting profile (black line) and the original unabsorbed profile (gray line). }
      
               \label{f1}
   \end{figure}
  
 An interesting feature of the models is the information provided by the profiles on the radial location of the dust within the ejecta. We show examples of changing the thickness and location of the dust containing region of the ejecta in Fig. 2.  If concentrated in the inner ejecta, the line asymmetry is pronounced but with no change in the maximum velocity.  The thicker the zone, or the more peripheral it is, the greater the obscuration of the outer ejecta and maximum observed recession velocity should decrease, while the blue wing of the profile remains virtually unchanged.   The effect of changing the inclination of the bicone  relative to the line of sight is shown in Fig. 3.  In each case, the top panel shows the predicted unobscured profile, the bottom shows the effect of the dust (at fixed radial location) for the same geometry.  We are neglecting the effects of scattering (see, however, Shore (2013) and the discussion, below).
  
  \begin{figure}
    \includegraphics[width=9.5cm]{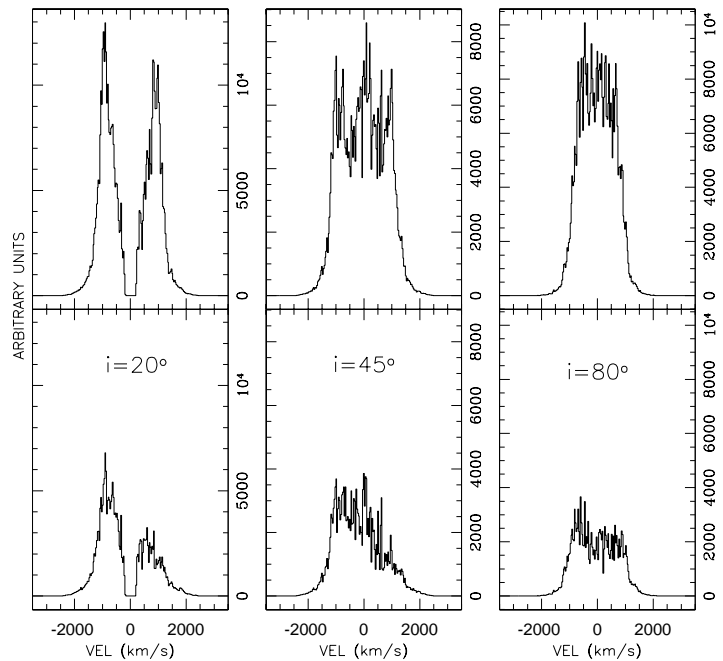}
      \caption{Effect of the dust within the ejecta as viewed from different inclinations. The first column show the unabsorbed line profile (top) and its absorbed version (bottom) for a biconical ejecta viewed at an inclination, $i$, of 20$^o$. The middle and right columns are as the left column but for inclination $i=45^o$ and 80$^o$. The adopted ejecta geometry in all cases is that of bicone having opening angle $\theta/2=60^o$ and thickness $\Delta r/R=0.6$. The dust is assumed to be in the inner half of each cone. }
               \label{f1}
   \end{figure}
 
 We now turn to the observations.  In the next sections we show how this approach can be applied to two recent classical novae that are known, from multiwavelength observations, to have formed dust during their outbursts.

\section{Observations and data reduction}

Nova V5668 Sgr = nova Sgr 2015b was first detected  by J. Search (CBET 4080) 2015 March 15.634 UT (JD 2457097.134, which we take as the zero point for dating) at a visual magnitude of around +6.   The optical light curve post-maximum ($m_V \approx 4$)  evinced a slow mean decrease while showing substantial, coherent excursions in magnitude over the next couple of months.    The light curve continued in this manner until Jun.1 when it entered the first stage of a major photometric decline of the sort experienced by DQ Her.  The visual band light curve is shown in Fig. 4 based on AAVSO photometry.   The multiwavelength view of the outburst has been presented by Gehrz et al. (2018), to which we refer the reader for more details.  Infrared observations have also  been described by Banerjee et al. (2016).  Here we concentrate on the implications of the high resolution line profile developments as they reveal the dust formation and evolution.

Our observational sequence covers the early stages and extends past the end of the dust event (see appendix A for the journal of observations) using  {\it HST}/STIS in the UV and with VLT/UVES in the optical.  The UVES optical spectra, originally taken under the DDT program 294.D-5051, were downloaded from the ESO archive. The observations consist of ten epochs between day 63 and 186 after outburst and cover the wavelength range $\sim$3000-9500 \AA\ at a resolution R$\sim$60000 (Table A1). 
The STIS spectrum was taken on day 235 after outburst (DDT program 14449). The HST visit was only a single orbit and used two different setups of the medium resolution (R$\sim$30000) echelle grating E230M to  cover the wavelength range 1200-2900\AA (Table A2).  All the data sets were reduced using each instrument pipeline.   The optical spectra were flux calibrated using the master response function built on a number of spectrophotometric standard stars observed in photometric conditions. Blue losses below the  Balmer series head occurred occasionally so the absolute calibration shortward of $\approx$3800\AA\ is unreliable. 

The nova was also observed with {\it Neil Gehrels Swift Observatory}/UVOT and  the UVOT monitored the nova between day 100 and day 420 after outburst in the broadband filters UVM2, UVW1, UVW2 (having central wavelength $\lambda_c$ 2246, 2600, and 1928 \AA, respectively). Low resolution grism spectra (R$\sim$75-100 and $\lambda$-range=1700 to 5000\AA) were taken between day 92 and 848 (see TableA3, A4).  The UVOT grism data reduction was done using Kuin  (2014) and Kuin et al. (2015) version 2.2, flagging bad spectral areas\footnote{see \url{http://www.mssl.ucl.ac.uk/www_astro/uvot/}}, and summing spectra from individual images to improve the signal to noise. For many spectra the brightest emission line cores exceeded the maximum brightness for recovering a flux calibration. The continuum and weaker lines were in the calibrated regime. Special care needs to be taken when interpreting the UVOT grism observations, see Appendix B for details.   The accuracy of the UVOT photometry depends on the source's brightness.   For sources with  brightness between  23rd and 13th magnitude, the photometry is properly extracted using aperture photometry on the images. For much brighter sources, up to about 9th magnitude, the readout streak that is  while the image data is moved to the readout buffer in a fraction of the frame time has been calibrated to give photometry up to about 9th magnitude. Even brighter sources also produce a readout streak, but their brightness shows inconsistent results with magnitude, see  Page et al. (2013).  

\section{V5668 Sgr through the dust event and its clearance}
      
The optical and UV light curves of V5668 Sgr show that the deep minimum began around 90 days after outburst when infrared continuum measurements (Fig. 4) indicated a rapid increase (ATel \#8753).  Until that time, the V magnitude had been decreasing only a few tenths of a magnitude per day.  Subsequently, the brightness rapidly decreased, reaching a minimum around day 120, followed by a slower recovery.  We (somewhat arbitrarily) take the end of the deep optical minimum  to be around day 170.         
    
 \begin{figure}
    \includegraphics[width=9.5cm]{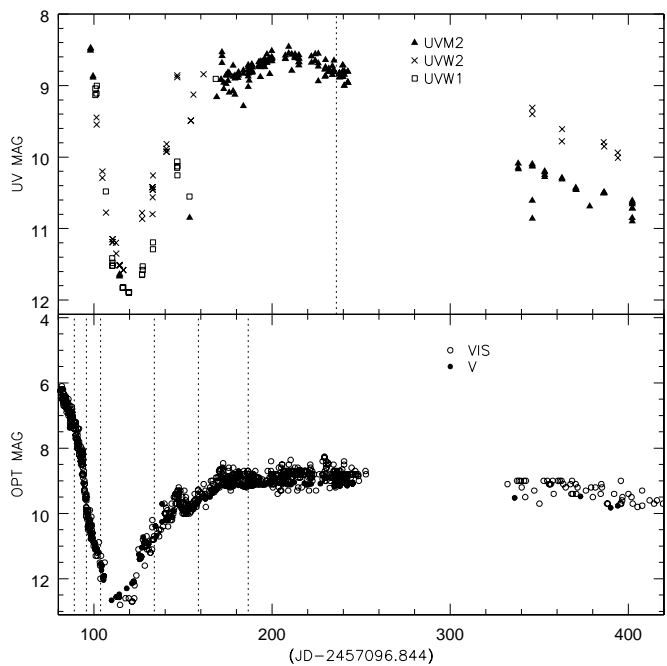}
      \caption{Top: UV light curve for V5668 Sgr (Swift) from day 80 through day 420,; the time of the HST/STIS spectrum is indicated by the  vertical line.  Bottom: visual (V and visual estimate) light curve from AAVSO data; the times of the UVES high resolution spectra are indicated by vertical lines. }     
         \label{f0}
   \end{figure}

 \subsection{V5668 Sgr: ionization dependent profile structure and extinction} 
 
 An important feature of the high resolution spectroscopic data is its ability to dissect the structure of the ejecta.  The  line profiles provide a picture of how the ionization changes within the ejecta because of the simplicity of the ballistic kinematics.  We start with an example of one of the most important set of lines, the [O I] 6300, 6364\AA\ doublet.  These are always optically thin and are ground state transitions.  Their ratio, when the density is sufficiently low for the ejecta to be considered in the nebular stage, is 3 (by the time of the displayed observations the [O I] 5577\AA\ line was no longer visible).  Thus, when the densities are low and the overlying absorption from the metallic lines is gone, the profiles trace the density structure.

 \begin{figure}
    \includegraphics[width=9.5cm]{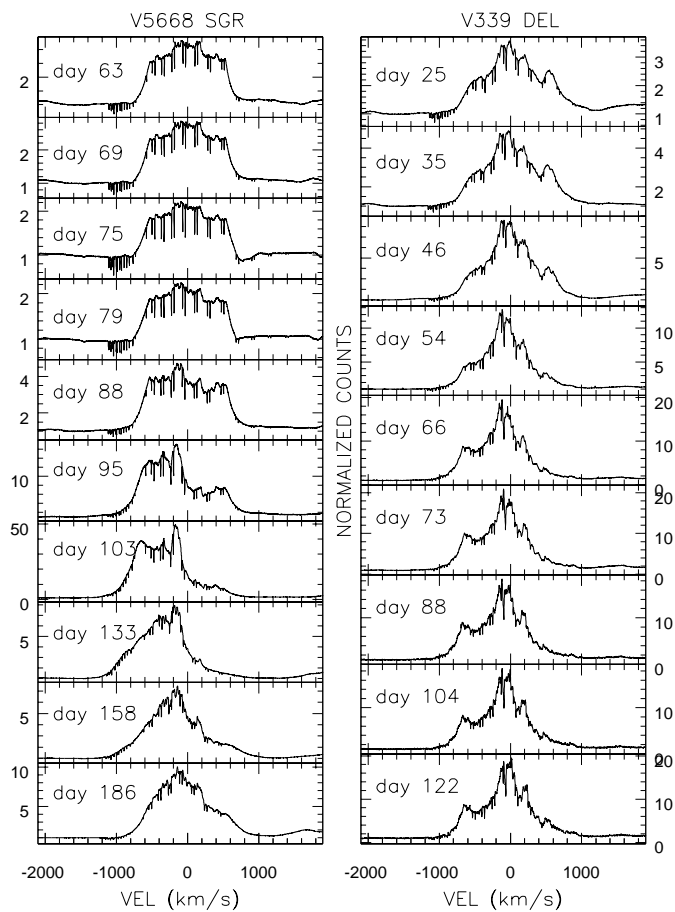}
      \caption{Comparison of the time development of [O I] 6300\AA\ for V5668 Sgr (left) and V339 Del (right).   An intrinsically optically thin transition, this forbidden line traces the effect of the dust formation event.  See text for details}
         \label{f0}
   \end{figure}
  
 We show an example of the time sequence for the [O I] 6300\AA\ line in Fig. 5.  The  panels show the ESO spectral sequence  during the dust events in V5668 Sgr (left) and V339 Del (right).    We concentrate, for the moment, on the V5668 Sgr sequence and will return to the comparative discussion in sec. 5.  The spectrum variations show clearly the disappearance of the redward part of the profile as early as around day 90, just prior to the onset of the deep {\it visual} photometric minimum.  Before that day, the profile was symmetric and the inferred $\tau_d$ was much smaller than unity.  We note that the extreme positive radial velocity of the wings does not change significantly despite the onset of line asymmetry.    We interpret this as the starting epoch for grain growth.   How the different ionization stages provide information about the process can be seen from a comparison of lines of two nitrogen ions, N$^+$ and N$^{3+}$, shown in Fig. 6.  We over-plot the model profiles for the two transitions using the optical depths derived from simulating the [O I] lines and the optical lines before the start of the photometric descent.   Resonance lines in the ultraviolet spectrum are uniquely suited to studying the structure of the ejecta and the V5668 Sgr spectra provide a well distributed set of  elements  manifesting a range of ionization stages.  In particular, nitrogen and carbon were present in three ionization stages, from singly to four times ionized.    In addition, recombination lines of a range of ions were present.   Table 1 presents the UV peak ratios measured on a broad range of ionization stage transitions for V5668 Sgr.  In all profiles, the blue to red peak ratio is nearly the same.   The dust therefore seems to have been obscuring the ejecta without a change in the ionization structure. We return to this point in sect. 6.

  
  \begin{table}
\flushleft\caption[]{V5668 Sgr: UV emission line blue-to-red peak ratios \textcolor{blue}{}}
\begin{center}
 \begin{tabular}{cccc}
 Ion & Wavelength (\AA) & Transition & Peak ratio \\
 \hline
 O V & 1371 & recombination & 1.16$\pm$0.05 \\
 N IV] & 1486 & resonance & 1.38$\pm$0.02 \\
  He II  & 1640 & recombination & 1.40$\pm$0.02 \\
  N IV] & 1718 &  recombination & 1.45$\pm$0.03 \\
  C III] & 1909 & resonance & 1.40$\pm$0.01 \\
  O IV & 2838 & recombination & 1.65$\pm$0.06 \\
  \end{tabular}
  \end{center}
  \end{table}
 
  \begin{figure}
    \includegraphics[width=8cm]{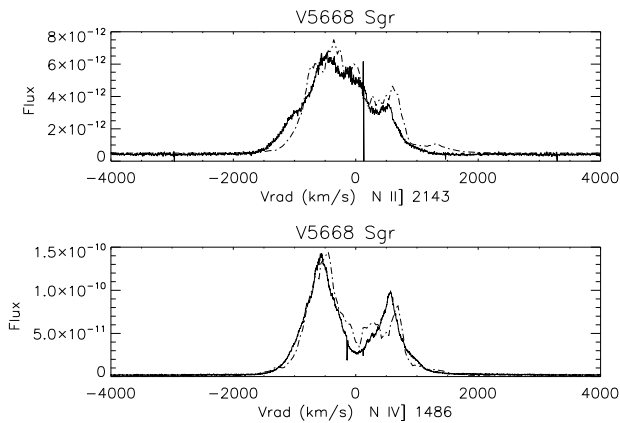}
      \caption{ Resonance lines of two nitrogen ions in the  V5668 Sgr {\it HST}/STIS observation compared to the dust model.   The model parameters are $\Delta r/R$=0.7, $v_{max}$=2500 km s$^{-1}$, $r_{in},r_{out}$=[0.35,0.5], $i=40^o$ and in the model units $\tau_d=1.3$.     Only the opening half-angle of the biconical ejecta differs between the model profiles: 40$^o$ for N$^+$ and 30$^o$ for N$^{3+}$.}
         \label{f2}
   \end{figure}

One way to check whether the line profile asymmetry is caused by density differences within the ejecta is to use the same procedure for obtaining the electron density that has been described in Mason et al. (2018).  We assumed the H$\gamma$ profile is the same as H$\beta$ on day 186, the last day for which we have high resolution ESO spectra.  Taking E(B-V)=0.3 from the Ly$\alpha$ and 21 cm H  I analysis (Gehrz et al. 2018), we scaled the H$\beta$ to subtract the contamination on [O III] 4363\AA.  The nebular line ratio, F(5007\AA)/F(4959\AA) $\approx$3, so the Balmer lines should have the same optically thin emission profiles.   Were the peaks due to density differences, we would expect the diagnostics to indicate a strong asymmetry with radial velocity.  In Fig. 7, we show this ratio and, even without deriving a specific value for the density, there is no strong asymmetry within the radial velocity range of the peaks.

\begin{figure}
    \includegraphics[width=7cm]{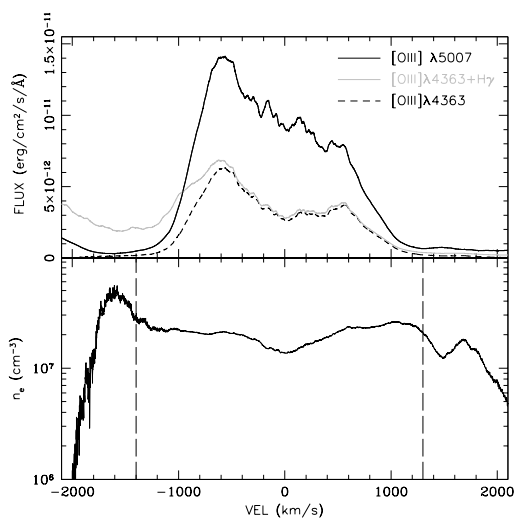}
      \caption{Electron density distribution in the V5668 Sgr ejecta for day 186 based on the optical [O III] profiles.  The 4363\AA\ is plotted without removing the H$\gamma$ contribution for visualization purposes.  The H$\beta$ line was used to model H$\gamma$, see text for discussion.     Notice that, like the nitrogen ratios, the derived $n_e$ is the same for the two lobes of the ejecta despite the profile asymmetries.}     
         \label{f0}
   \end{figure}
 
 For V5668 Sgr we have the advantage of a high resolution infrared spectrum at approximately the same time as the STIS observations that includes the He II 1.96$\mu$ line.\footnote{We sincerely thank Prof. Banerjee for providing these data.  See also Gehrz et al. (2018) for additional line profiles.}  The IR spectrum was taken on  2016 Feb. 28 at Mt. Abu with a resolution of $\approx$1000.   We compare the STIS He II 1640\AA\ with that profile in Fig. 8.  Notice that the IR line peak ratio is 1.10$\pm$0.05 while at about the same time the UV line shows a ratio about 60\% greater.  This further supports our contention that such profile asymmetries are not the result of density differences in the two lobes of the ejecta.

\begin{figure}
\begin{center}
    \includegraphics[width=7cm]{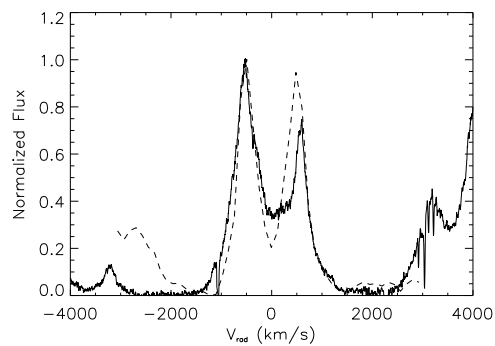}
      \caption{Comparison of the [Si VI] 1.96$\mu$ profile from day 350 (dash) with He II 1640\AA\ from the STIS spectrum (solid).  This highlights the difference in the extinction between the infrared and UV.  See text for details.} 
      \end{center}
         \label{f0}
   \end{figure} 
 
           \subsection{Extinction law and dust properties}
       
The wavelength dependence of the dust extinction is provided by comparing the asymmetries of the UV and optical profiles of the He II  lines.  Since the 1640\AA, 4686\AA\ and 5411\AA\  lines arise from recombination, their ratio has been used like the Balmer lines to estimate interstellar extinction.  We can, however, extend this by noting that the two sides of the emission profile arise in different parts of the ejecta and, therefore, sustain differential {\it internal} extinction. The overall effect of any intervening interstellar medium toward the source is not important for the relative asymmetries.  Guided by the simulations, we can take the flux ratio between two peaks to provide the optical depth  $\tau_d \approx \pi a^2 C_{abs}(\lambda) N_d$, where $C_{abs}$ is the absorption efficiency (Draine \& Lee 1984), $a$ is the grain radius,  and N$_d$ is the dust column density.  Since the geometry is obtained independently from modeling the overall emission line profiles, the dust {\it volumetric} number density can be estimated using the maximum velocity and the time since outburst  (e.g. Mason et al. 2018).  

The infrared has a far lower dust optical depth than the visible or ultraviolet, the well exploited advantage that during the Fe-curtain stage of the outburst, so  that spectral range provides a view through the ejecta at a time when the shorter wavelengths are obscured (e.g., Hayward et al. 1996; Evans \& Gehrz 2012). The same profile structures that appear elsewhere appear in the infrared at an earlier epoch than at shorter wavelengths. This advantage is reversed for the post-dust forming stage.  For the standard interstellar extinction law, for instance, the absorption ratio is $\sim$12 at He II 4686\AA\ and $>$700 for He II 1640\AA.  The choice of wavelengths is for comparison between profiles of a single species.  Thus, even when the infrared is virtually transparent, the ultraviolet can continue to show the effects of differential obscuration for a much longer time.  This is  what the V339 Del sequence and the V5668 Sgr single epoch data show.  The UV profiles for V339 Del are much more asymmetric than the IR lines at the same date after outburst.  Furthermore, the  structure evident from the comparison of ionized species, such as He II and N IV, is more asymmetric than simultaneously observed Balmer and optical [O I] lines.

A  quantitative constraint on the extinction law for the dust can therefore be obtained from the profiles of ultraviolet and optical He II recombination lines that are widely spaced in wavelength.  For instance, we compared the 5411\AA\ (day 186) and 1640\AA\ lines (day 235) (blending caused difficulties for the He II 4686\AA\ profile so we opted for the weaker line in this case).  If the dust extinction had a wavelength dependence like the interstellar law, with $\tau_d \approx 1$ derived from the optical line using the blue to red peak ratio (about 1.4), that on the ultraviolet line would be about 2.  Instead, the two values are nearly equal, around 1.4, an indication that the dust is grey.  This also agrees with the lack of a strong enhanced absorption at around 2200\AA.  While this latter behavior could result from silicate grains, the same extinction law (grey and no bump) results even if the dust is carbonaceous (Draine \& Lee 1984) provided the grain radius is $\gtrsim$0.1$\mu$.  A similar conclusion was reached from the UV variations of V705 Cas (Shore et al. 1994).   We also note that besides the infrared coronal [Si VI] line, all other lines discussed in the late time IR spectra of V5668 Sgr were almost perfectly symmetric (Gehrz et al. 2018) at only 100 days after the STIS spectrum,  lending further support to dust -- not density -- as the cause of the profile structure at shorter wavelengths.  

Scattering in aspherical ejecta has a secondary effect on  the visible continuum, depending on $C_{sca}$. Since the peak ratio yields information only about $C_{abs}$,  the infrared emissivity rise at the onset of dust formation may be accompanied by a color tendency that makes the continuum appear relatively bluer.  Even if the wavelength dependence of the absorption is relatively flat, the scattering may increase toward shorter wavelengths.  This depends on the size spectrum of the grains and the relative angle of viewing the ejecta.   Another effect of the scattering depends on where the dust forms in the ejecta, as outlined in Shore (2013).   During the outburst of novae having the type of light curve called ``C''  (cusp) by  Strope et al. (2010), emission lines may increase in intensity and broaden when the dust forms if the dust has a scattering coefficient comparable to that for absorption.  Draine \& Lee (1984) show that large grains behave this way.

The dust mass can be estimated using the radius $R(t)$, a choice of the possible grain mass density (we use the Evans et al. (2017) estimate from their study of V339 Del, $\rho_d = 2.2$ g cm$^{-3}$), and only geometric properties of the ejecta.  At day 250, the radius was 10$^{15}$ cm, thus giving a column  density for the grains of about $10^{-6}$cm$^{-2}$, so the derived dust mass is a few times $10^{-9}$M$_\odot$.   This is  smaller than the derived dust mass in Gehrz et al. (2018) but the grain sizes are also quite different in that work, by about a factor of ten (the study used 2$\mu$ grains for the maximum grain size) and it is based on an assumed distance to derive the infrared luminosity.  Accounting for that difference raises our mass by a factor of three.  We note, however, that the calculations of dust {\it mass}, as opposed to column density, depend on the assumed geometry and the published estimates assume sphericity.  Our derived mass is actually nearly the same as the CO mass spectroscopically derived in Banerjee et al. (2016).  We remark that our spectroscopic estimate is independent of any distance estimates to the nova or luminosity considerations while any value based on intensity requires a distance.  Tracing the optical depth back to the photometric minimum and assuming a constant dust to gas ratio, $\tau_d \sim t^{-2}$ gives an optical depth of around 6 for day 105, consistent with the value $\approx$ 7.5 cited by Gehrz et al. (2018) and with the depth of the minimum obtained from the spectroscopic sequence (Fig. 5).  It thus appears that the grains survived long after the end of the photometric event.   Since the emissivity of the dust is expected to have $\beta \approx 2$ for Mie scattering in the infrared, the dust temperature at 250 days should scale as $T_d(t) \sim (R(t)/R(t_0))^{-1/3} \approx 0.7 T_d(t_0)$, so if at IR maximum the temperature was about 1000K, at times later than around day 250 it should be around 700 K for optically thin emission.  This is consistent with the reported infrared continuum measurements (Gehrz et al. 2018).    
        
\subsection{Continuum variations during the dust event}  

Our grism coverage for V5668 Sgr is similar to the low resolution 1200-3000\AA\ spectroscopic sequence obtained with {\it IUE} (Shore 1994) for V705 Cas.  In addition, we have the UVES flux calibrated  optical spectral sequence that was unavailable in the V705 Cas coverage.   The two sequences are shown in Fig. 9, keyed to the optical light curve (shown at right on the two panels).  The top panel shows the UV grism sequence, the bottom  displays the contemporaneous sequence of binned optical spectra.  At the start of the dust event, the ultraviolet was at the end of the Fe-curtain stage, which complicates the interpretation.  The absorption bands were changing along with the overall continuum level as the Fe-curtain became progressively more transparent.  The net decrease is, however, approximately uniform in the mean level across the entire spectrum.  The uniformity of the extinction is more clearly seen in the lower panel, where we show the UVES sequence.  The changes in continuum level are obvious  and wavelength independent (i.e., grey).   Note, for instance,  that the O I lines at 7773\AA\ and 8446\AA\ vary only in absolute, not relative, intensity during this descent in brightness.

 \begin{figure*}
    \includegraphics[angle=270,width=14cm]{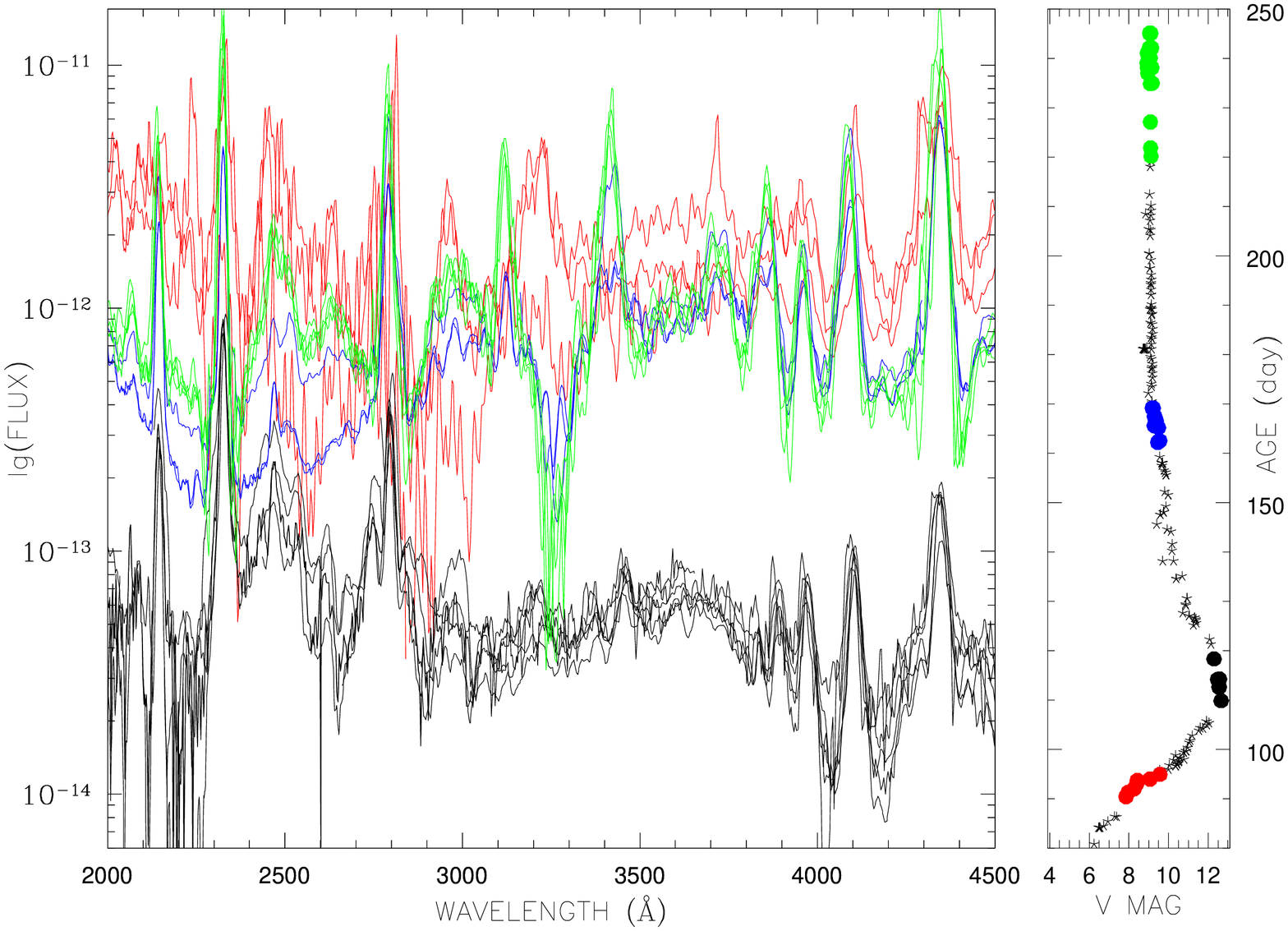}

    \includegraphics[angle=270,width=14cm]{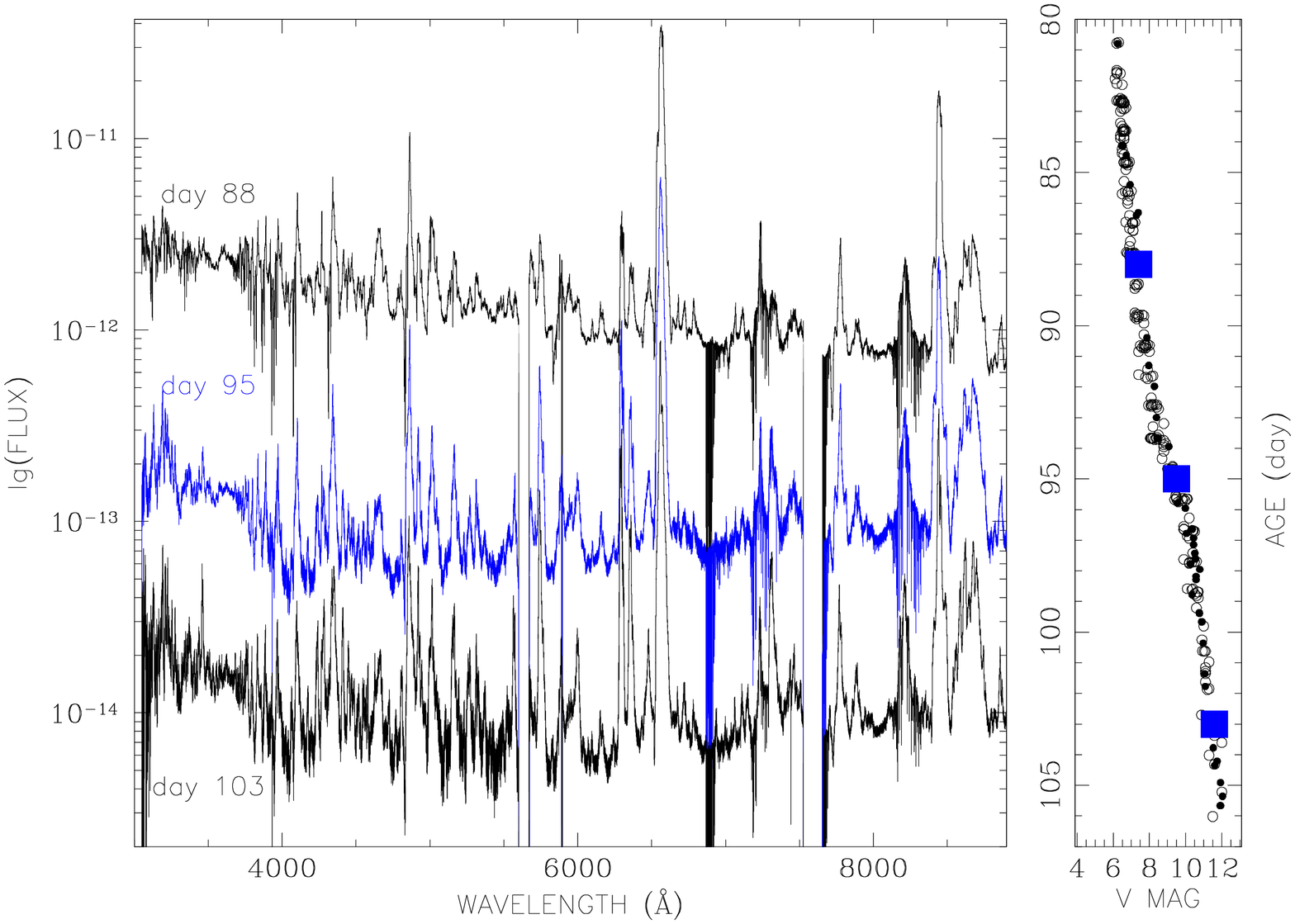}
      \caption{Top: {\it Swift} grism spectra spanning the dust  event for V5668 Sgr, from day 80 through day 420.  The colors refer to the stage indicated in the light curve  (right, note that latest date is at the top of the light curve column).     Bottom: UVES/VLT spectra spanning the dust forming interval for V5668 Sgr, from day 80 through day 110. The epochs are indicated by the blue squares on the light curve (right, note that the light curve is displayed with latest date at the bottom, the curves align with the photometry).  See text for more details. }     
         \label{spc_ev}
   \end{figure*}

       \section{A tale of two novae: comparison of V5668 Sgr with V339 Del}
    
 In this section, we compare the {\it STIS}  spectrum of V5668 Sgr with V339 Del, another CO nova observed with good cadence and very high resolution in the UV by {\it STIS}, as described in Shore et al. (2016).  Our original intention was to have a baseline for comparison to see what might be the differences in their basic properties since the spectra are of comparable quality and precision.  We show some representative line profiles in Fig. 10.  The N IV] 1486\AA, aside from being a recombination doublet, is also a resonance transition while He II 1640\AA\ is from recombination.  To say they are close seems an understatement.  This cannot be mere coincidence, it bespeaks a common underlying mechanism.   
 
 \begin{figure}
 \begin{center}
    \includegraphics[width=9.0cm]{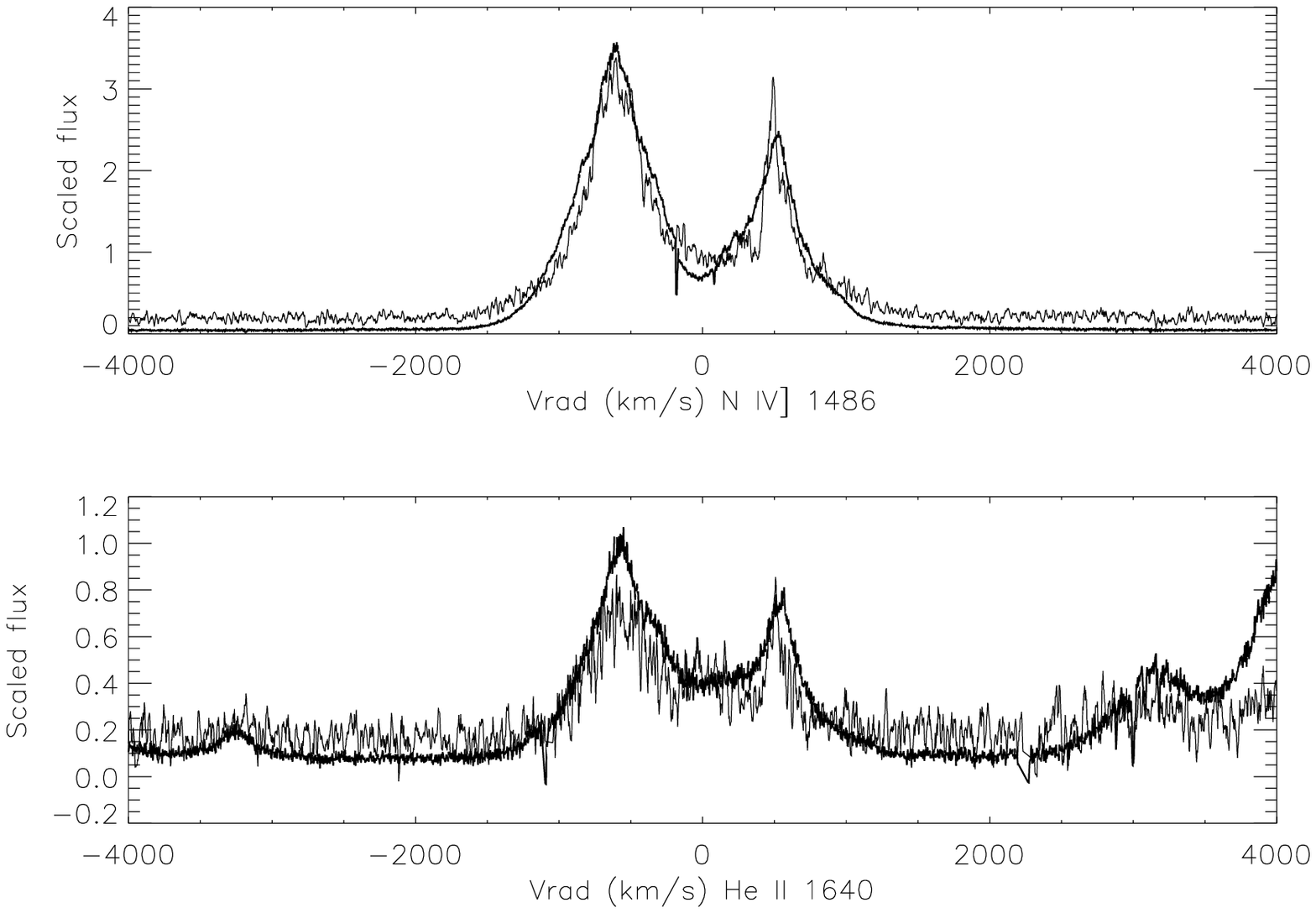}
      \caption{Scaled comparison of the N IV] and He II ultraviolet lines from the STIS spectra of V5668 Sgr (thick line) and V339 Del (thin line, day 242).   Five point boxcar smoothing has been applied to the V339 Del spectrum but without rectification of the profiles.}     
      \end{center}
         \label{f0}
   \end{figure}
   
Since, for ballistic motion, the form of the profile is most sensitive to the overall symmetry of the ejecta, the comparison indicates either that the structure of the ejecta are nearly the same for two completely unrelated objects or that they share a ``third parameter'' besides abundance and geometry.  That property appears to be that both formed dust.    This is supported by the UV light curve of V339 Del, shown in Fig. 11, that shows a far greater decrease in the UV at about day 77 than was seen in the optical photometry.  The difference in the photometric behavior of these two novae can easily be accounted for assuming different inclinations and  geometries.  The covering factor for V5668 Sgr was clearly the larger,  the inner ejecta and central star were almost completely obscured at the maximum of the dust optical depth while for V339 they were more visible.   The same conclusion is drawn from the [O I] sequence in Fig. 5.   Note that the onset of the profile asymmetry was a bit before day 54, before the UV descent.  This could mean that the obscuration of the inner ejecta on the central source was substantially lower than for V5668 Sgr but the lobes of the ejecta were still self-obscured as the dust began forming.  Even if the cone parameters are similar they are not the same and it suffices to incline V339 Del a bit more to the line of sight to considerably diminish the maximum obscuration.  The rest of the photometric evolution, as the dust clears, should be virtually the same.    The IR continues to radiate as the radiation temperature drops and, taking the simplest hypothesis that the grain properties do not change, the line profiles will also change similarly.    
   
 \begin{figure}
    \includegraphics[width=9.5cm]{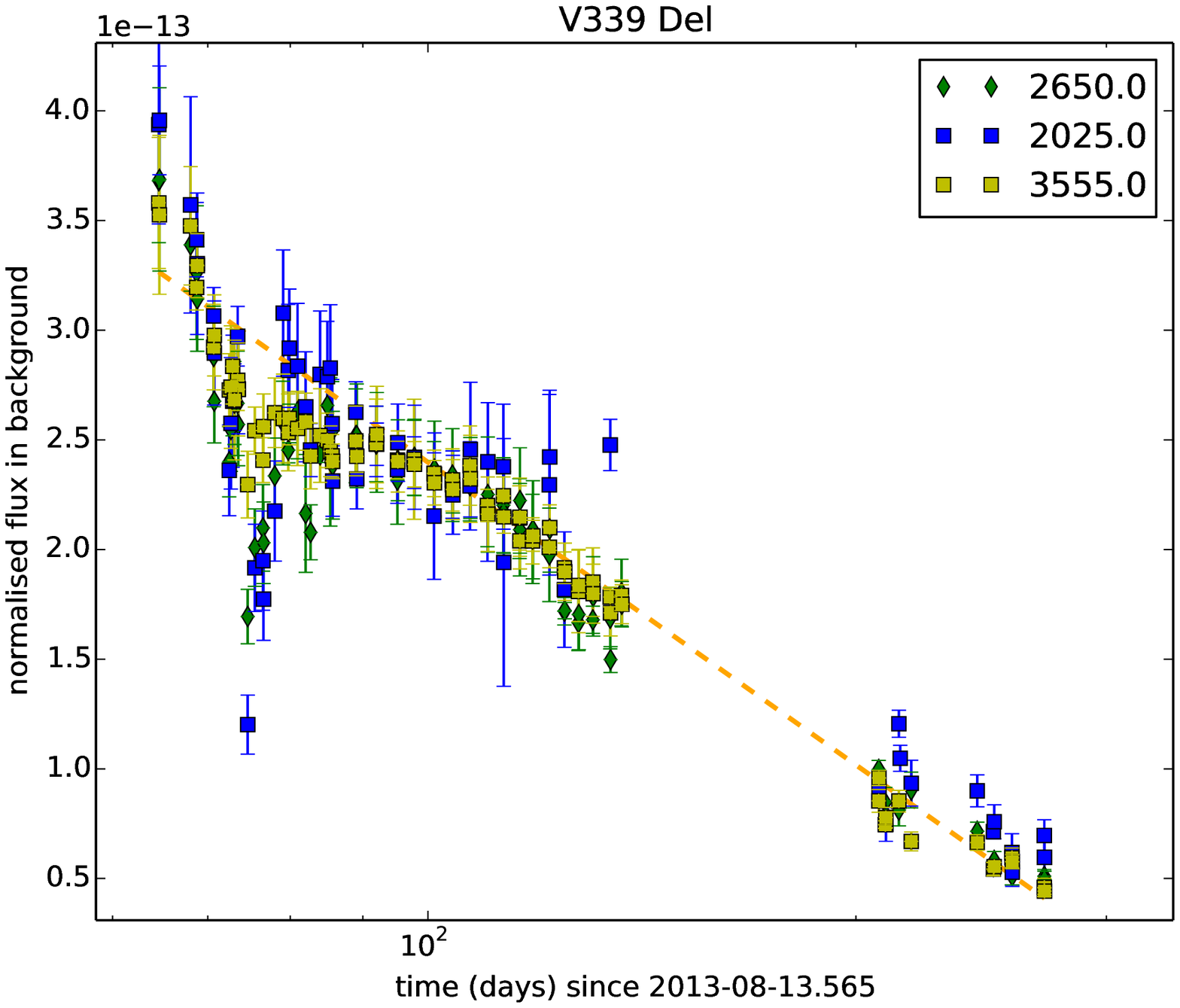}
      \caption{UV light curve for V339 Del (Swift) from day 80 through day 420. }     
         \label{f0}
   \end{figure}

  \subsection{Infrared development compared}
  
The infrared view of the V339 Del dust forming event was presented by Evans et al. (2017).  We will concentrate on their discussion of the line profile evolution, since this was followed with sufficient resolution to compare with optical data of the event.  The Paschen series line profiles were initially dominated by the P Cyg absorption component of the optically thick ejecta until around day 50.  Thereafter, as Evans et al. show in their Fig. 5, once the line  absorption had disappeared from the blue side of the profiles, the lines became shifted toward the blue and asymmetric with a suppressed red wing compared to the earlier stages.  They briefly mention the possibility that the line shape may have been altered by dust obscuration but do not further explore its implications.  The profiles then recovered symmetry, appeared double peaked, and  remained the same thereafter.   The authors interpret this as evidence of grain destruction by exposure to the hard radiation coming from the central source.  
  
Based on the IR alone this is a consistent picture but, when combined with other wavelength intervals, the spectroscopic development leads to a different conclusion.  In particular, our late optical and UV observations of V339 Del (from day 635 to 866, see Mason et al. 2018) systematically showed asymmetric line profiles that evinces the survival of the grains well after the Evans et al. claim.   The spectra from day 650 (Fig. 13) require $\tau_d\approx 1$, corresponding to a dust mass estimate of $6\times 10^{-10}$M$_\odot$ for $\approx 0.1\mu$ grains.  Evans et al. (2017) derive a dust mass of about $5\times 10^{-9}$M$_\odot$ but use a larger grain radius and underestimate the rate of expansion of the ejecta.  The ratio of the two estimates of optical depth scales approximately $v_{max}^3a^2(\Omega/4\pi)$,  where $\Omega$ is the total solid angle of the bicone (see Shore et al. (2016) for further discussion of the modeling).  Note that the published mass estimate assumes spherical ejecta.   Given the difference in assumed parameter values,  our estimate should be a factor of $\approx 3$ {\it lower} than the published value.

A check on the greyness of the dust is provided by comparing  our UV and optical He II lines  from around day 640 with fig. 5 of Evans et al. (2017); the infrared helium and Paschen line profiles were symmetric on day 683.  This is consistent with a minimal difference in density of the bicones and that the grains were {\it still} present in the expanding material, instead of being destroyed earlier by photon and collisional processes.  Banerjee et al. (2016) derived a dust mass for V5668 Sgr of about $3\times 10^{-7}$M$_\odot$ based on the infrared energy distribution.  V5668 Sgr had a strong molecular precursor stage of CO (no other molecules were reported)  with a proposed  near equality for  $^{12}$C relative to $^{13}$C based on the emission bandheads in the near IR.  Gehrz et al. (2018) found a similar mass for the dust as Banerjee et al. (2016), assuming a mean density of about 2 g cm$^{-3}$, and no indication of the silicate emission features.  This is also consistent with the CO nova identification.

 \begin{figure}
 \begin{center}
    \includegraphics[width=9.0cm]{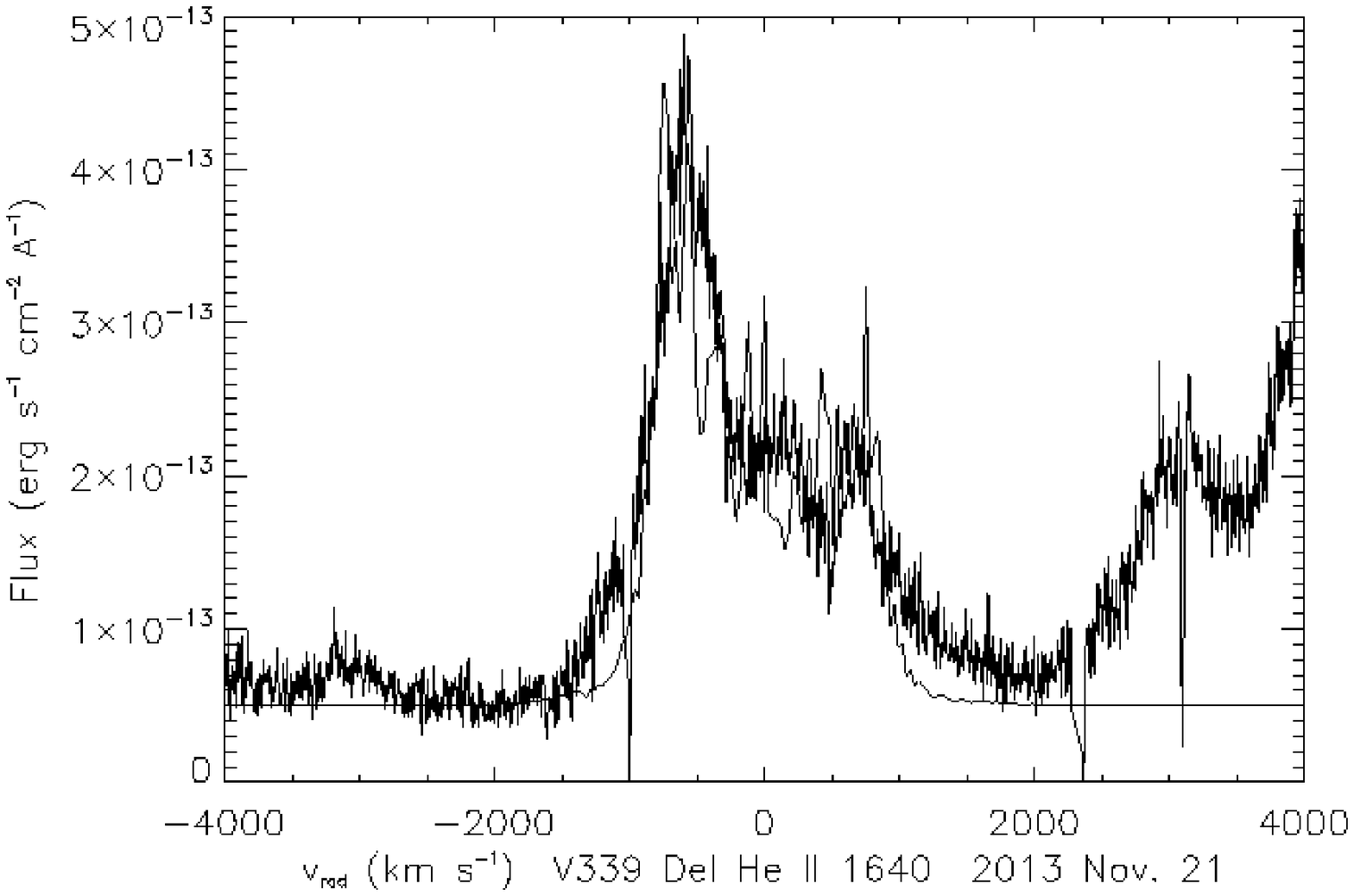}
      \caption{Comparison of model and He II 1640\AA\ profile for V339 Del from the HST/STIS spectrum of 2013 Nov. 21 (day 80, following the dust forming event).  The optical depth was around unity.}
         \label{f0}
         \end{center}
   \end{figure}

The  comparison of the optical and UV He II lines for V339 Del, results in dust properties that  are similar to V5668 Sgr.  Using the high resolution profiles presented in Shore et al. (2016) and Mason et al. (2018), the red to blue peak ratios show only small differences between the two wavelength intervals.   For the He II lines on day 88, the peak ratio was 2.4 for 1640\AA\ and $\ge$1.9 for 4686\AA.  The latter was blended so this is a lower limit based on extrapolating the wing .   For the last  spectra for the purpose, from 2015 May, the He II 4686\AA\ line shows a peak ratio of 1.1$\pm$0.1 that is the same as He II 1640\AA.  These last profiles are compared in Fig. 14.  This was around Day 650, that corresponds to a late time IR detection reported by Evans et al. (2017).  The He II profiles were more asymmetric than those of V5668 Sgr, which may indicate a difference in the internal distribution of the dust (the peak ratios depend more on $\tau_d$ than the radial shell within the ejecta and on the ionization structure, see Fig. 2).  
The depth of the UV photometric minimum was also smaller than V5668 Sgr by about a factor of 20,  $3 < \tau_{3000,max} < 4$ at around day 80, as we show in Figures 11 and 12.  The optical depth on day 88 is consistent with $\tau_{d,1640} \approx 2$, so that at the start of the dust forming episode the optical depth in the UV was about 2.8, which is also consistent with a factor of 10 to 20 decrease in the photometry.  Ballistically extrapolating forward to around day 650 assuming no intrinsic changes, the predicted optical depth at 1640\AA\ is about  0.3.   This  is about a factor of 3  smaller than the value implied by the models for that last epoch, $\tau_d \approx 0.9$ but in the opposite sense with respect to expectations of grain destruction.  Although the explanation for this failure remains open at present, it seems that the overall development of the two novae was quite similar.  The grains had a wavelength independent opacity, like those in V5668 Sgr, and  {\it the dust optical depth at maximum was also large despite the rather insignificant glitch in the visual light curve}.

The implication of this last result is interesting in that the minimum for V339 Del was so much briefer in duration and shallower in magnitude than either V705 Cas or V5668 Sgr.  One possibility is that the mass of dust formed was significantly lower in V339 Del than the ``deep dippers''.   Interestingly, V339 Del did not show any infrared CO emission during the optically thick Fe curtain stage, although the atomic carbon lines were strong.  The inferred N/C value is consistent with expectations of a TNR yield ratio for a CO nova (Mason et al. 2018) based on late stage absorption lines from the ejecta.  Additionally, although we do not have absolute abundances for the individual elements, the [Ne IV] 1602\AA\ and [Ne V] 1575\AA\ lines were certainly absent, supporting this subclass identification.    Thus, overall, there was nothing particularly deviant about the white dwarf or the abundances in V339 Del compared with other CO novae.  The onset of the dust formation, signaled by both the UV photometric descent and infrared rise, was early but not exceptionally so.     The shell geometries and dynamics off the two novae were not that different yet the dust mass yields seem to differ by as much as a factor of 10.  The comparison thus points to there being a range of dust  yields in classical nova ejecta from the same abundance class of the progenitor white dwarf.

\begin{figure}
    \includegraphics[width=9.5cm]{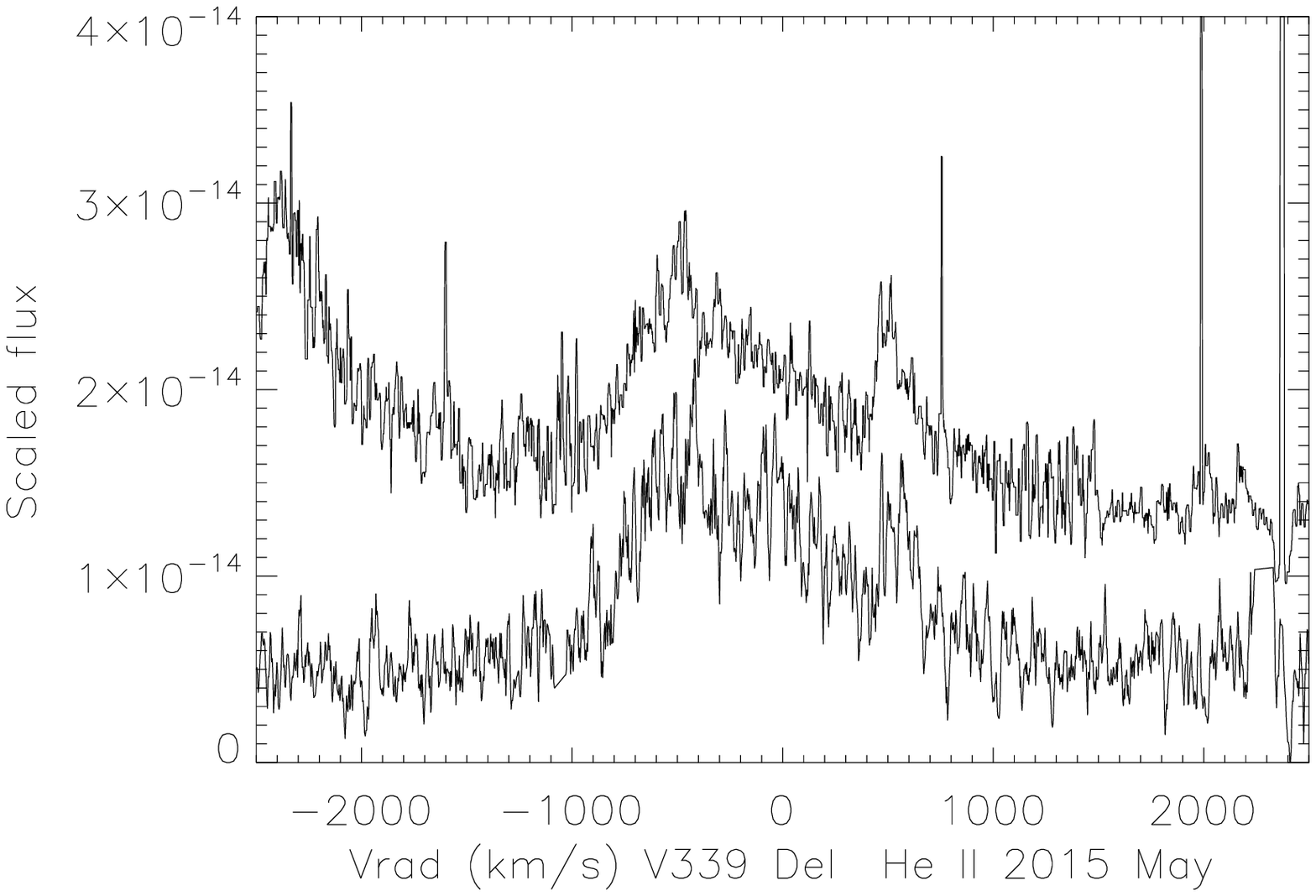}
      \caption{V339 Del: comparison of He II 1640\AA\ (bottom) to He II 4686\AA\ (top) in 2015 May.  This was the last contemporaneous UV-optical pair available for comparing the dust properties.  The optical spectrum was seven point median filtered, fluxes are scaled.  See text for discussion. }     
         \label{f0}
   \end{figure}

\section{Dust signatures in other novae}

For classical novae, the possibility of some sort of grain production was first raised in the 1930s, connected with the early observations and interpretation of DQ Her 1934.  This nova has become the phenomenological prototype of a specific photometric behavior among novae; Stratton \& Manning (1939), Payne-Gaposchkin \& Whipple (1939), Payne-Gaposchkin (1957), McLaughlin (1960), and  Beer (1974) provided extensive summaries of the various stages of the spectral development.  In particular, Payne-Gaposchkin (1957) wrote ``{\it The bright lines, especially the forbidden lines had been saddle-shaped for some time. Although they persisted during the drop in brightness they all decreased greatly in intensity, and they redward maxima all but disappear}''.   In the Stratton \& Manning photographic atlas, the [OI] line  shows the onset and increase of the asymmetry prior to Apr. 25, the last date presented.  All reports of the DQ Her event  highlighted the changes in the optical emission line profiles just before and during the deep minimum that began about 60 days after outburst.    With hindsight,  in light of these studies, it has been inferred that T Aur 1891 was another such event (e.g., Clerke 1903, Payne-Gaposchkin 1957).  The  deep dust forming event was observed  when, around 1892 Mar. 7, the nova began a steep decline from about a visual magnitude of 5.6 to below 12 mag by Mar. 28.    Vogel (1892) and Campbell (1892)  report changes in the symmetry of the line profiles although these are more subjective than for DQ Her and later novae.

In the CCD era, the best studied deep minimum dust former was V705 Cas 1993.  This chanced to occur during the lifetime of the {\it IUE} satellite and the sequence of low resolution spectra spanning the onset of the dust event was presented in Shore et al. (1994).  The event was also covered extensively in the infrared  using spectrophotometry and photometry (Evans et al. 2017).   The grism data for V5668 Sgr is comparable, as we noted, but because there is no published high resolution profile study of V705 Cas we cannot further compare the ejecta structure or details of the dust formation so we pass to the next historical example.  From the 1200-3000\AA\ region alone and using only low resolution ($\approx$300) spectrophotometry during the event, the derived dust absorption was grey.

Published line profiles  for V723 Cas (Evans et al. 2003) show the usual biconical double peak at day 200, but only infrared lines were shown.  Optical spectra presented by Goranskij et al. (2007) from around day 200 are symmetric and similar to the infrared profiles,  indicating an optically thin, biconical structure.  Based on the spectra, there appears to be no signature of dust in the ejecta.  This nova, by all accounts, was exceptional in being an extremely slowly developing event.  Nevertheless, it would appear that not all CO novae form dust, or some do so in sufficient quantity that the emission lines are not strongly affected or are formed over a more extended zone of the ejecta than that harboring dust. 

The peculiar case of V2362 Cyg 2006 actually fits well into this overall picture of dust formation and evolution.  Lynch et al. (2008) presented a comprehensive summary of the photometric and spectroscopic development of the outburst.  This nova displayed one of the anomalous types of light curves that shows a secondary maximum, at around the same time as the DQ Her-type dippers show a deep minimum, with the characteristic infrared signature of dust formation.  At no time, however, did the emission lines show any asymmetry.  Instead, at the maximum of the re-brightening event, they significantly broadened to about the same width as in the optically thick stage but without associated P Cyg-type absorption.  Our argument regarding the dust is essentially the same for this nova except that the bicones were oriented such that they were at very large inclination to the line of sight (Shore 2013) and the broadening is the result of {\it scattering} with about the same efficiency as absorption.   

Some indication of the effect may be present in the spectra shown for V809 Cep by Munari et al. (2014); in their Fig. 11 the day 220 and day 352 profiles of [O III] 5007\AA\ shows the asymmetry we have described and this is a nova for which dust formation was also inferred from the infrared.   The nova V1425 Aql is a  counter-example of the notion that {\it all} CO novae are dust formers.  Infrared observations were, however, obtained only 600 days after outburst, more than 200 days after the inferred XR turnoff and when the source was too faint in the UV for further observations (Lyke et al. 2001).   Line profiles shown in Evans \& Gehrz (2012) are consistent with the later, optically thin infrared lines of different ionization stages displayed by V339 Del (Evans et al. 2017).  Had there been an infrared component at that time it would have been too weak .  

The same spectroscopic signature was observed in the recent dust forming CO nova V1324 Sco (Finzell et al. 2018).  Although the line profiles presented in that paper cover only the period before the main event, that part of the expansion when the ejecta transitioned to optically thin in the Fe-curtain, one Balmer line profile taken 800 days after the deep minimum shows precisely the same signature as we have shown here for late spectra.  What had been a symmetric profile when the line absorption vanished was blue-asymmetric.  Although there were no spectra taken during the deep minimum with sufficient resolution to be useful, the persistence of the dust signature nearly two years after the event supports our contention that the dust is formed during the minimum and survives thereafter.  More to the point, as Sakon et al. (2016) demonstrate for V1280 Sco, the dust is still detected spectrophotometrically even 2000 days after outburst.

We finally come to V1369 Cen.  In our earlier analysis (Mason et al. 2018) we concentrated on the spectroscopic indicators of geometry and dynamics.  But we noted that a peculiar asymmetry appeared in the profiles even in the late time spectra that was not evident in the plasma diagnostics.  The similarity of the late time profiles to those of V5668 Sgr and V339 Del suggest that, possibly, here too some dust was formed in the outburst.  From the comparison of the STIS spectra from 2014 Nov. 3 and 2014 Mar. 7, the later one is more symmetric although the peak ratio is 1.0$\pm$0.02.  The N IV] 1486\AA\ line is more indicative, the ratio in the earlier spectrum was 1.20 while in the second it is indistinguishable from unity.   We also note that the AAVSO photometry for this nova shows a brief minimum of about $\Delta V \approx 1$ mag (about the same for B) around days 80 to 100, similar to most dust forming novae.   We note, however, that this nova showed an anomalously large N/C excess relative to solar and, perhaps, this plays some role in  dust formation.  While neither the profile modeling nor the photometric variations is a ``smoking gun'', we suggest that it is consistent with such an event and, as such, argues that grain formation in classical novae is a more common feature than photometry alone would indicate.

 \section{Discussion}

There post-visual maximum increase in infrared continuum emission rising toward longer wavelengths and longward of around 2$\mu$ is the one, unambiguous, signature of the appearance of grains in the ejecta: (see, e.g., Evans \& Gehrz 2012).  This  broadband emission can include the SiO and SiC emission features, but these may not be present depending on the dust properties.   The decrease in optical brightness, the ``deep minimum'' characteristic of the DQ Her-type events, may not happen if the dust laden parts of the ejecta do not obscure a sufficient solid angle of the pseudo-photosphere.  In addition, the line profiles will always have some random asymmetries (Mason et al. 2018) resulting from the explosive, single ejection nature of the outburst, but those will not be of a unique type or change in the systematic way we have discussed.  

The infrared has so low a dust opacity relative to the UV that the lines can return to their early symmetric form in a comparatively short time; the continuum emission will be observable for far longer.  The flux should, however, decrease in time because of dilution and (depending on the details of the grains, i.e., their composition and size), the temperature should steadily decrease.  Since the structures are frozen within the ejecta, any change in IR continuum emission should be a simple power law in time.  This suggests that the ejecta from dust formers may be visible in millimeter continuum photometry and imaging for far longer than would be inferred from the disappearance of the IR line profile asymmetries.  Infrared emission from optically thin illuminated grains will persist as long as the grains survive and the central WD remains luminous.  The turn-off times for XR emission have been derived from the cessation of the SSS stage (Schwarz et al. 2011) or from the onset of expansion-controlled recombination (e.g., Vanlandingham et al. 2001, Shore et al. 2014).  Infrared dust emission will, however, still be powered by the unobservable EUV and FUV emission before the WD returns to quiescence.  Thus, if we assume a constant grain radius and wavelength dependent absorption efficiency $Q_{\rm eff} \sim \lambda^{-\beta}$, where $\beta \approx$ 2, then the grain temperature should scale as $T_{IR} \sim t^{-2/(\beta+4)}$.  For V339 Del, the dust temperature at day 36 was around 1600K, around day 650, it was inferred to be about 650 K (Evans et al. 2017); we would expect $T_{IR} \sim$ 600K.  In other words, the grains likely survived the XR phase and the emission continued until the central source turns off.  This is consistent with the persistence of the line asymmetries in the UV and optical (De Gennaro Aquino et al. 2015, Shore et al. 2016).  For V5668 Sgr, the  infrared emission remained detectable even after the appearance of infrared coronal lines (Banerjee et al. 2016).  Note that the photodestruction and collision rates decrease  faster than the dust optical depth so the high ionization is not necessarily an indication that the grains are gone.

Among the ONe novae for which we have high resolution observations at sufficiently long times, V959 Mon (Shore et al. 2014), 
 V382 Vel (Della Valle et al. 2002, Shore et al. 2003), LMC 2000 (Shore et al. 2003), and V1974 Cyg (Shore et al. 1993) were obtained at around day 150, none shows the asymmetries found in the CO novae.  None shows any anomalous event in its light curve of the sort observed in the three CO novae.  While there is little infrared data available for any of the ONe novae, the later time nebular spectra are consistent with only optically thin gas without a grain component.   V838 Her was a clear outlier in having displayed a weak infrared excess that has been interpreted as a dust event, but this was by all measures an extreme outburst (e.g., Schwarz et al. 2007).  It may be mere coincidence, but given the similarities of the ONe together (e.g., Shore et al. 2013), it seems likely that there is an essential difference between the two abundance classes regarding dust yields.   We conjecture that dust formation is not only common in the CO novae, something that has been known for a long time, but that neither the recurrent novae (like T Pyx and U Sco) nor the ONe group produce significant amounts of dust, if any.   This implies that the presolar grains attributed to novae are from the CO group, hence their isotopic abundance patterns should be a fossil record of the nucleosynthesis in that subgroup that cannot be otherwise teased out of the spectroscopy, for example, Amari \& Loders (2007) and Illiadis et al. 2018).     

To continue this point, the scenario of dust survival also aligns with evidence for a nova contributor to the presolar grain samples.  Recall that even ``classical'' novae presumably recur on long timescales.  Take this timescale to be about $10^4$ yrs (a popular choice based on the hibernation picture).  With a few dust forming novae per year in the Galaxy, each contributing about the same amount, we  expect to have accumulated  a few solar masses of nova enriched grains in the course of Galactic evolution.  Consequently, novae are not major players in the dust budget but they have a unique nucleosynthetic signature (e.g., Casanova et al. 2016).

\section{Conclusions}

In summary, we find that there is no need to invoke dust destruction at late times in either V5668 Sgr or V339 Del, even under irradiation by the X-ray and EUV from the central star.  It appears that there was little change in the dust properties over relatively long times after the photometric minimum.   In addition, the evidence from the UV and optical spectrophotometric variations for both novae, like those of V705 Cas, is that the dust was {\it not} present before the start of the steep decline, that it grew rapidly, and that the event approximately coincided with the beginning of transparency in the ultraviolet.   After the opacity maximum, the dust absorption optical depth merely decreases through expansion-driven dilution. The grains are large, with radii of at least 0.1$\mu$.  For V5668 Sgr, the optical and UV profiles should become increasingly symmetric; at around 800 days after outburst,  the dust optical depth will have fallen to about $\tau_d \lesssim$0.1.  There are no observations yet from such late stages of the expansion.  The grains would, however, continue to emit even at this late time if the central source is still powering the emission.  Assuming, again for argument's sake, that the white dwarf bolometric luminosity remains constant for a very extended interval.  The temperature should still be about 500 K, after almost three years, which {\it ceteris paribus} is an upper limit.  It would be of considerable interest to attempt an ALMA observation of the dust at late stages at sub-mm and mm wavelength.   

This scenario of dust survival also aligns with evidence from pre-solar grains, for which a nova contribution has been identified based on isotopic ratios (e.g., Amari \& Loders 2007; Illiadis et al. 2018).  Recall that even ``classical'' novae presumably recur on long timescales.  Again, for argument's sake, take this timescale to be about $10^4$ yrs (a popular choice based on the hibernation picture).  With a few dust forming novae per year in the Galaxy, each contributing about the same amount, we  expect to have accumulated  a few solar masses of nova enriched grains in the course of Galactic evolution.  Novae are, consequently, is not major players in the dust budget but they have a unique nucleosynthetic signature (e.g., Casanova et al. 2016).

We remark in closing that our survey of the literature highlights a serious lacuna: the rarity of published high resolution optical and infrared line profiles for dust forming novae.  Most of the discussions have been based on low resolution flux measurements and spectral energy distributions.  There is also very little spectropolarimetry at sufficient resolution to study the three dimensional structure of the ejecta across line profiles.   Nevertheless, as we have discussed, there are a number of comprehensive spectroscopic studies in the last few decades that  provide evidence in support of our hypothesis.   We contend that the similarity of the spectroscopically and photometrically derived dust masses and properties should encourage observers to put more effort into obtaining the necessary data.  The reward will be a significant extension in our understanding of the dust properties and evolution in novae.

\begin{acknowledgements}
Based on observations made with the Nordic Optical Telescope, operated by the Nordic Optical Telescope Scientific Association at the Observatorio del Roque de los Muchachos, La Palma, Spain, of the Instituto de Astrofisica de Canarias. The research leading to these results has received funding from the European Union Seventh Framework Programme (FP7/2007-2013) under grant agreement No. 312430 (OPTICON).  Based on observations made with the NASA/ESA Hubble Space Telescope, obtained at the Space Telescope Science Institute, which is operated by the Association of Universities for Research in Astronomy, Inc., under NASA contract NAS 5-26555. These observations are associated with program \# 13828.   Support for program \#13828 was provided by NASA through a grant from the Space Telescope Science Institute, which is operated by the Association of Universities for Research in Astronomy, Inc., under NASA contract NAS 5-26555 and ESA HST DDT proposal 14449.  NPMK was supported under a grant from the UK Space Agency. 
We acknowledge the useful discussions within the Swift Nova-CV community, the program was supported by the 
Swift PI, Neil Gehrels, who supported this research with the Swift resources. We thank the Swift planners and operators for the extra work required with the grism offsets.  We sincerely thank Prof. D.P.K. Banerjee for providing the late time infrared spectrum of V5668 Sgr.  We thank  Kim Page, Jordi Jos\'e, Greg Schwarz, Bob Gehrz, Nye Evans, and Bob Williams for discussions and exchanges, and  SNS thanks the Astronomical Institute of the Charles University for a visiting appointment during which this work was begun.
\end{acknowledgements}

\newpage

\appendix

\section{Journal of Observations: V5668 Sgr}

\begin{table*}
	\centering
	\caption{V5568 Sgr: UVES log of observations.}
	\label{tbl1}
	\begin{tabular}{lcccccc} 
		\hline
		UT date & day & \multicolumn{4}{c}{exptime (s)} &  \\
		  (start)       & (since outburst) & 346 & 564 & 437 & 760 &  \\
		\hline
 2015-06-03 & 80 & 60,400 & 3$\times$3,7$\times$9  & 5$\times$3,100 & 5$\times$3,10$\times$3  &  \\
 2015-06-12 & 89 & 60,400 & 3$\times$3,7$\times$9 & 5$\times$3,60 & 5$\times$3,10$\times$2 &  \\
 2015-06-19 & 95 & 602$\times$2 & 15$\times$23 & 5$\times$3,60$\times$2 & 5$\times$3,15$\times$2 &  \\
 2015-06-27 & 103 & 600$\times$3 & 20$\times$10,100$\times$6 & 100,200 & 100,200 &  \\
 2015-07-27 & 134 & 100,570$\times$2 & 100,660$\times$2 & 100,550$\times$2 & 100,550$\times$2 &  \\
 2015-08-20 & 158 & 50$\times$4,100$\times$6,180$\times$6 & 50$\times$4,100$\times$6,180$\times$6 & 50,100,180$\times$3 & 50,100,180$\times$3 &  \\
 2015-09-17 & 186 & 50$\times$3,100$\times$3,200$\times$3 & 50$\times$3,100$\times$3,200$\times$3 & 50$\times$3,100$\times$3,180$\times$3 & 50$\times$3,100$\times$3,180$\times$3 &  \\
		\hline
	\end{tabular}
\end{table*}

 \begin{table*}
\flushleft
 \caption[]{V5668 Sgr: HST/STIS log of observations. \textcolor{blue}{}}
 \label{log}
\begin{tabular}{lcccc}
		 \hline
obs-date (start) & day & epoch &setup & exp  (s) \\
		 \hline
2015-11-6.07  & 235 & late-SSS & E230M/1978 & 575 \\
2015-11-6.08   & 235 & late-SSS   & E230M/2707 & 575 \\
		 \hline
\end{tabular}
\end{table*}

\begin{table*}
	\centering
	\caption{V5568 Sgr: Swift UVOT Grism log of observations used.}
	\label{uvotlog}
	\begin{tabular}{lcccccc} 
		\hline
  mid-date/time (UT)& day(s) &duration(s)&  filter & \# exposures\\ 
\hline
   2015-06-16T06:43 &   92.6 &  289& UG&1 \\  
   2015-06-17T14:57 &   94.0 & 1062& UG &1\\  
   2015-06-18T22:53 &   95.3 & 1043& UG &1\\   
   2015-06-20T22:37 &   97.3 &  997& UG &1\\   
   2015-06-21T08:34 &   97.7 & 5533& UG &2\\                
   2015-06-22T17:50 &   99.1 & 6255& UG &2\\               
   2015-06-24T03:25 &  100.5 &64493& UG+VG &3\\               
   2015-06-26T20:58 &  103.2 &  881& UG &1\\ 
   2015-06-28T00:06 &  104.4 & 1050& UG &1\\ 
   2015-06-30T01:20 &  106.4 & 1588& UG &1\\  
   2015-07-03T14:10 &  110.0 &17651& UG+VG &3\\             
   2015-07-05T15:28 &  112.0 &13083& UG &3\\                
   2015-07-07T15:31 &  114.0 &12302& UG &3\\               
   2015-07-09T12:00 &  115.9 &30163& UG+VG &3\\              
   2015-07-12T09:01 &  118.7 &46490& UG+VG &2\\                
   2015-07-20T08:17 &  126.7 &24327& UG+VG &3\\             
   2015-07-26T04:39 &  132.6 &23578& UG+VG &3\\               
   2015-08-02T23:26 &  140.3 &13146& UG+VG &3\\               
   2015-08-09T00:53 &  146.4 & 6692& UG &2\\ 
   2015-08-16T16:17 &  154.0 & 1587& UG &1\\  
   2015-08-18T03:25 &  155.5 &  890& UG &1\\   
   2015-08-23T12:42 &  160.9 & 6420& UG &2\\            
   2015-08-23T20:42 &  161.2 &  649& UG &1\\ 
   2015-08-30T15:24 &  168.0 & 6932& UG &2\\               
   2015-09-03T08:46 &  171.7 &  901& UG &1\\  
   2015-09-06T10:13 &  174.8 &  889& UG &1\\  
   2015-09-09T07:06 &  177.7 &  889& UG &1\\  
   2015-09-12T11:33 &  180.8 &  889& UG &1\\   
   2015-09-16T11:28 &  184.8 &  270& UG &1\\      
   2015-09-20T06:15 &  188.6 &  889& UG &1\\ 
   2015-09-23T22:09 &  192.3 &  889& UG &1\\ 
   2015-09-27T20:14 &  196.2 &  884& UG &1\\   
   2015-10-10T16:14 &  209.0 & 1170& UG &1\\  
   2015-10-25T00:54 &  223.4 &  973& UG &1\\   
   2015-10-29T08:37 &  227.7 &  914& UG &1\\   
   2015-11-04T00:15 &  233.4 &  653& UG &1\\   
   2015-11-06T22:29 &  236.3 &  850& UG &1\\ 
   2015-11-13T17:14 &  243.1 &  742& UG &1\\  
   2016-02-24T08:55 &  345.7 & 6511& UG &2\\  
   2016-03-12T01:23 &  362.4 & 6178& UG &2\\ 
   2016-04-04T12:29 &  385.9 &18056& UG &2\\ 
   2016-04-12T10:10 &  393.8 & 6239& UG &2\\ 
   2017-07-10T15:32 &  848   & 1054& UG &1 \\ 
		\hline
	\end{tabular}
\end{table*}

\begin{table*}
	\centering
	\caption{V339 Del: Swift UVOT Grism log of observations used.}
	\label{uvotlog}
	\begin{tabular}{lcccccc} 
		\hline
  mid-date/time (UT)& day(s) &duration(s)&  filter & \# exposures\\ 
\hline
   2013-10-21T22:53 &   68   & 112567 & UG & 4 \\
   2013-10-23T05:54 &   70.7 &  901.3 & UG & 1 \\ 
   2013-10-25T05:55 &   72.7 & 1062.7 & UG & 1 \\
   2013-10-26T02:45 &   73.6 &  946.6 & UG & 1 \\
   2013-10-27T05:59 &   74.7 &  955.5 & UG & 1\\
   2013-10-28T02:49 &   75.6 & 1062.7 & UG & 1\\
		\hline
	\end{tabular}
\end{table*}

\newpage

\section{Description of the peculiarities in the $Swift$ UVOT Grism spectra}

These notes are specific for the spectra taken of Nova V5668 Sgr and V339 Del 
presented here. A general discussion of Swift UVOT grism data can be found in 
Kuin et al. 2015. The spectra which have been displayed in Fig. 9  have 
not been cleaned of contaminating zeroth orders and other defects described below.   

The early grism observations of V5668 Sgr were made using a Swift ToO which restricted 
the observation to be done without an offset on the detector. Observations thus done
are placed with the target in the center of the field of view, and the spectrum falls 
across the center of the detector. We used to option where the filter wheel is rotated 
across the aperture, "clocked" in Swift talk, which masks part of the zeroth orders in 
the field. Following the insight of Keith Mason, a change was made to the way the grism 
was mounted in the XMM-OM and the grism was mounted such that the first order in this 
clocked configuration falls on the detector where the zeroth orders are masked away. 
This is the configuration we used for both the UV and Visible grisms (named UG and VG in
the tables). 

The nova was initially too bright for the UVOT detector which uses a photon counting 
detector consisting of a phosphor screen, 3-stage MCP, fibre-optic, and a CCD where 
centroiding of the photon splash gives the photon position. For very high flux the MCP 
recharge time limits the measurement, and similarly, the CCD frame time is 11 ms, so 
at high fluxes the probability of multiple photons incident in one CCD frame is high. 
This coincidence loss can be corrected statistically by measuring enough frames. 

These characteristics have some effect on the spectra. The MCP threshold is not reached 
for the grism first order. The early time spectra from day 92-97.3 show in the continuum 
evidence of coincidence loss that is too high to correct, and the loss peaks where the 
grism throughput is highest, around 2700~$\AA$ in the UV grism. The continuum from day 
97.7 on is consistent with the photometry and shows no dip around the peak sensitivity. 
However, the line emission is much larger than the continuum. The center of the lines that
are formed in the 2000-4000~$\AA$ region are all too bright for a good coincidence loss 
correction, except during the faintest part of the minimum and at very late times. 
After the dip in brightness, from day 168-230, the coincidence loss once again affect 
the line cores. 

In the first spectra, which were taken at the center of the detector, the second order 
also falls across the first order, starting at 2750~$\AA$, and the second order response 
peaks in the UV. However, for the V5668 Sgr spectra, the bright lines of N~III] and C~III]
at 1750 and 1909~$\AA$ are also very bright in the second order spectrum. These show up
as extra peaks in the first order around 2950~$\AA$ and 3220~$\AA$, respectively. 
If the spectra are at a certain offset on the detector then the second order falls next 
to the first order, tracing it nearly parallel but merging at higher wavelengths. 
Now the N~III] and C~III] lines do not fall on top of the first order spectrum, but they 
turn out to be so strong still, that they create a coincidence loss pattern around them, 
stealing photons away from the first order. Due to the offset of the second order the 
effect of photon loss in the first order is shifted to slightly longer wavelengths. This 
can be seen in the later spectra during the brighter period as an extra pair of dips in 
the first order around 3010 and 3250~$\AA$. The effect of the other second order lines 
is smaller and difficult to see. The most prominent is the 2325(2) line that falls 
near 4300~$\AA$. Unfortunately, mistakes get made with the offset requests, and we 
see therefore a varying amount of these competing effects in the spectra. 

The faintest spectra become noisy because of the background noise, which in the clocked 
observing modes is restricted mostly below 2300~$\AA$. At a flux of $1.0 10^{-14}$ erg 
cm$^{-2}$ s$^{-1}$ the S/N level is close to 1. 

In the V339~Del spectra, some spectra suffered from first orders near the nova 
spectrum caused by other stars in the field. When an unwanted first order falls over 
the nova spectrum, just a few photons more can affect the spectrum where the 
grism sensitivity is low, since there the conversion from count rate to flux is the 
largest. This leads to a flux spectrum showing an upturn in the UV or the yellow/red part 
of the spectrum.


\begin{thebibliography}{}
\bibitem[Corrales(2015)]{2015ApJ...805..23C} Amari, S. \&  Lodders, K. 2007, HiA, 14, 349
\bibitem[Corrales(2015)]{2015ApJ...805..23C} Banerjee, D.P.K., Joshi, V., Srivastava, M. \& N. M. Ashok, N.M. 2016, ATel 8753
\bibitem[Corrales(2015)]{2015ApJ...805..23C} Banerjee, D. P. K., Srivastava, M. K., Ashok, N. M., \& Venkataraman, V. 2016, MNRAS, 455, L.109
\bibitem[Corrales(2015)]{2015ApJ...805..23C} Beer, A. 1974,, Vistas Astr., 16, 179
\bibitem[Corrales(2015)]{2015ApJ...805..23C} Bevan, A. \& Barlow, M.J. 2016, MNRAS, 456, 1269
\bibitem[Corrales(2015)]{2015ApJ...805..23C} Campbell, W.W. 1892, PASP, 4, 231
\bibitem[Corrales(2015)]{2015ApJ...805..23C} Casanova, J., Jos\'e, J., Garcia-Berro, E., \& Shore, S. N. 2016, A\&A, 595, A28
\bibitem[Corrales(2015)]{2015ApJ...805..23C} Clerke, A. 1903, {\it Problems in Astrophysics} (London: Adam \& Charles Black)
\bibitem[Corrales(2015)]{2015ApJ...805..23C} De Gennaro Aquino, I., Schr\"oder, K.-P., Mittag, M., Wolter, U. et al. 2015, A\&A, 581, 134
\bibitem[Corrales(2015)]{2015ApJ...805..23C} De Gennaro Aquino, I., Shore, S. N., Schwarz, G. J., Mason, E. et al. 2014, A\&A, 562, A28 
\bibitem[Corrales(2015)]{2015ApJ...805..23C} Della Valle, M., Pasquini, L., Daou, D. \& Williams, R. E 2002, A\&A, 390, 155
\bibitem[Corrales(2015)]{2015ApJ...805..23C} Derdzinski, A. M.; Metzger, B. D., \& Lazzati, D. 2017, MNRAS, 469, 1314
\bibitem[Corrales(2015)]{2015ApJ...805..23C} Draine, B. T. \& Lee, H. M. 1984, ApJ, 285, 89
\bibitem[Corrales(2015)]{2015ApJ...805..23C} Evans, A. \& Gehrz, R.D. 2012, BASI, 40, 213
\bibitem[Corrales(2015)]{2015ApJ...805..23C} Evans, A., Banerjee, D. P. K., Gehrz, R. D., Joshi, V. et al. 2017, MNRAS, 466, 4221
\bibitem[Corrales(2015)]{2015ApJ...805..23C} Finzell, T., Chomiuk, L., Metzger, B. D., Walter, F. M. et al. 2018, ApJ, 852, 108
\bibitem[Corrales(2015)]{2015ApJ...805..23C} Gehrz, R. D., Evans, A., Helton, L. A., Shenoy, D. P., Banerjee, D. P. K. et al. 2015, ApJ, 812, 132
\bibitem[Corrales(2015)]{2015ApJ...805..23C} Gehrz, R.D., Evans, N., Woodward, C.E., Helton,  L.A., Banerjee, D.P.K. et al. 2018, ApJ, in press (arXiv:180400575)
\bibitem[Corrales(2015)]{2015ApJ...805..23C} Goranskij, V. P., Katysheva, N. A., Kusakin, A. V., Metlova, N. V. t al. 2007, AstBu, 62, 125
\bibitem[Corrales(2015)]{2015ApJ...805..23C} Hayward, T.L.,  Saizar, P., Gehrz, R. D., Benjamin, R. A. et al. 1996, ApJ, 469, 854
\bibitem[Corrales(2015)]{2015ApJ...805..23C} Iliadis, C., Downen, L. N, Jos\'e, Jordi, Nittler, L. R. \& Starrfield, S. 2018, ApJ, 855, 76
\bibitem[Corrales(2015)]{2015ApJ...805..23C} Kuin, N.~P.~M., 2014, Astrophysics Source Code Library, record ascl:1410.004
\bibitem[Corrales(2015)]{2015ApJ...805..23C} Kuin N.~P.~M., Landsman, W., Breeveld, A.~A., et al., 2015, MNRAS 449, 2514
\bibitem[Corrales(2015)]{2015ApJ...805..23C} Lucy, L. B., Danziger, I. J., Gouiffes, C., \& Bouchet  1989, LNP, 350, 164
\bibitem[Corrales(2015)]{2015ApJ...805..23C} Lyke, J. E., Gehrz, R. D., Woodward, C. E., Barlow, M. J. et al. 2001, AJ, 122, 3305
\bibitem[Corrales(2015)]{2015ApJ...805..23C} Lynch, D.K., Woodward, C.E., Gehrz, R.D., Helton, L.A. et al. 2008, AJ, 136, 1815
\bibitem[Corrales(2015)]{2015ApJ...805..23C} McLaughlin, D. B. 1937, POMic, 6, 107
\bibitem[Corrales(2015)]{2015ApJ...805..23C} Mason, E., Shore, S.N., De Gennaro Aquino, I., Izzo, L., Page, K., and Schwarz, G. J. 2018, ApJ, 853, 27
\bibitem[Corrales(2015)]{2015ApJ...805..23C} Munari, U., Maitan, A., Moretti, S. \& Tomaselli, S. 2015, NewA, 40, 28
 \bibitem[Corrales(2015)]{2015ApJ...805..23C} Munari, U.,  Ochner, P., Dallaporta, S., Valisa, P. et al. 2014, MNRAS, 440, 3420 
 \bibitem[Corrales(2015)]{2015ApJ...805..23C} Page, M.~J., Kuin, N.~P.~M., Breeveld, A.~A., et al., 2013, MNRAS, 436, 1684
\bibitem[Corrales(2015)]{2015ApJ...805..23C} Payne-Gaposchkin, C. 1957, {\it The Galactic Novae} (Amsterdam: North-Holland) 
\bibitem[Corrales(2015)]{2015ApJ...805..23C} Pozzo, M.;,Meikle, W. P. S., Fassia, A., et al. 2004, MNRAS, 352, 457
\bibitem[Corrales(2015)]{2015ApJ...805..23C} Payne-Gaposchkin, C. \& Whipple, F.  1939, Har.Cir 433
\bibitem[Corrales(2015)]{2015ApJ...805..23C} Rawlings, J.M.C. 1988, MNRAS, 232, 507
\bibitem[Corrales(2015)]{2015ApJ...805..23C} Sakon, I.,  Sako, S., Onaka, T., Nozawa, T. et al. 2016, ApJ,  817, 145
\bibitem[Corrales(2015)]{2015ApJ...805..23C} Schwarz, G. J., Ness, J-U, Osborne, J. P., Page, K. L., et al. 2011, ApJS, 197, 31
\bibitem[Corrales(2015)]{2015ApJ...805..23C} Schwarz, G. J., Shore, S. N., Starrfield, S.,\& Vanlandingham, K. M. 2007, ApJ, 657, 453
\bibitem[Corrales(2015)]{2015ApJ...805..23C} Shore, S. N., Schwarz, G., Bond, H. E., Downes, R. A. et al. 2003, AJ, 125, 1507
\bibitem[Corrales(2015)]{2015ApJ...805..23C} Shore, S. N.,Starrfield, S., Gonzalez-Riestrat, R., Hauschildt, P. H., \&  Sonneborn, G. 1994, Natur, 369, 539
\bibitem[Corrales(2015)]{2015ApJ...805..23C} Shore, S. N. 2012, BASI, 40, 185
\bibitem[Corrales(2015)]{2015ApJ...805..23C} Shore  2013, A\&A, 559, L7
\bibitem[Corrales(2015)]{2015ApJ...805..23C} Shore, S. N., Mason, E., Schwarz, G. J., Teyssier, F. M., et al. 2016, A\&A, 590, A123
\bibitem[Corrales(2015)]{2015ApJ...805..23C} Shore, S. N., De Gennaro Aquino, I., Schwarz, G. J., Augusteijn, T., et al.  2013, A\&A, 553, A123 
\bibitem[Corrales(2015)]{2015ApJ...805..23C} Shore, S. N. \& Gehrz, R. D.  2004, A\&A, 417, 695
\bibitem[Corrales(2015)]{2015ApJ...805..23C} Shore, S. N., Sonneborn, G., Starrfield, S., Riestra-Gonzalez, R., \&  Ake, T. B 1993, AJ, 106, 2408
\bibitem[Corrales(2015)]{2015ApJ...805..23C} Starrfield, S.; Iliadis, C., Timmes, F. X., Hix, W. R. et al. 2012, BASI, 40, 419
\bibitem[Corrales(2015)]{2015ApJ...805..23C} Stratton, F. J. M. \& Manning, W. H. 1939, {\it Atlas of spectra of Nova Herculis 1934} (Cambridge: Solar Physics Observatory)
\bibitem[Corrales(2015)]{2015ApJ...805..23C} Strope, R. J., Schaefer, B. E., \&  Henden, Arne A. 2010, AJ, 140, 34
 \bibitem[Corrales(2015)]{2015ApJ...805..23C} Vanlandingham,  K. M., Schwarz, G. J., Shore, S. N., \& Starrfield, S. 2001, AJ, 121, 1126
\bibitem[Corrales(2015)]{2015ApJ...805..23C} Vogel, H. 1893, Abhand. K\'onlig. Akad.  Wissen. (Berlin), 1, 115, A108
\end{thebibliography}
\end{document}